\documentclass[12pt]{article}
\usepackage[margin=1in]{geometry}
\usepackage{setspace, indentfirst}
\everymath{\displaystyle}
\usepackage{amsmath, amssymb, mathtools}

\usepackage{enumitem}

\usepackage{footmisc}

\usepackage{float, makecell, caption, multicol, multirow, booktabs, threeparttable}
\usepackage{xcolor,graphicx}
\newcommand{\refitem}{\noindent\hangindent=2em\hangafter=1}

\title{Automatic Order, Bandwidth Selection and \\ Flaws of Eigen Adjustment in HAC Estimation}
\author{Zhuoxun Li, Clifford M. Hurvich}
\date{}
\begin{document}
\maketitle
\begin{abstract}
    In this paper, we propose a new heteroskedasticity and autocorrelation consistent covariance matrix estimator based on the prewhitened kernel estimator and a localized leave-one-out frequency domain cross-validation (FDCV). We adapt the cross-validated log likelihood (CVLL) function to simultaneously select the order of the prewhitening vector autoregression (VAR) and the bandwidth. The prewhitening VAR is estimated by the Burg method without eigen adjustment as we find the eigen adjustment rule of Andrews and Monahan (1992) can be triggered unnecessarily and harmfully when regressors have nonzero mean. Through Monte Carlo simulations and three empirical examples, we illustrate the flaws of eigen adjustment and the reliability of our method.
\end{abstract}
\newpage

\section{Introduction}
In econometrics, considerable attention has been paid to develop heteroskedasticity and autocorrelation consistent (HAC) covariance matrix estimators. The key quantity to estimate is the long-run variance (LRV) of the product of regressors $X_t$ and errors $u_t$
\begin{equation*}
    S_V = \sum_{r=-\infty}^\infty\Gamma_V(r), \quad V_t = u_tX_t \in \mathbb{R}^{d+1} \ \ \text{and} \ \ E(V_t) = 0, \quad t = 0,\dots,n-1,
\end{equation*}
where $\Gamma_V(r) = E(V_tV_{t-r}')$ is the autocovariance of $V_t$ at lag $r$. If $\{V_t\}$ is weakly stationary with absolutely summable autocovariance function, then it has spectral density function
\begin{equation*}
     f_V(\omega) = \frac{1}{2\pi}\sum_{r=-\infty}^\infty \Gamma_V(r)e^{-i\omega r},
\end{equation*}
where $\omega \in [-\pi, \pi]$ denotes the frequency and $S_V = 2\pi f_V(0)$.

Therefore, any LRV estimator can be viewed either as an estimate of the infinite sum of autocovariances from a time domain perspective or as an estimate of the spectral density function at zero from a frequency domain perspective. Frequency domain cross-validation has not been widely used in the HAC problem\footnote{Since LRV estimation is the core problem in HAC estimation, these two terms are used interchangeably.} because it traditionally focuses on estimating the spectral density function over a set of frequencies, not at a single frequency zero that the HAC problem concerns. Note that since $u_t$ is unobserved in practice, LRV estimators are usually constructed on $\hat{V}_t = \hat{u}_tX_t$, where $\hat{u}_t$ denotes the estimated error.

Newey and West (1987) first proposed the kernel estimator, a weighted sum of sample autocovariances of $\hat{V}_t$ from lag $-m$ to $m$. The parameter $m$ is called the bandwidth and how to choose $m$ is where two perspectives differ (a detailed discussion is given in section IV). It turns out that kernel estimators do not perform well when the spectral density function of $\{\hat{V}_t\}$ is not smooth around zero. To solve this problem, Andrews and Monahan (1992) proposed the prewhitened kernel estimator. First, they estimate a VAR($q$) (usually $q=1$) model on $\{\hat{V}_t\}$ by OLS and extract the residuals $\{\widetilde{V}_t\}$:
\begin{equation}
    \hat{V}_t = \sum_{i=1}^q\hat{A}_i\hat{V}_{t-i} + \widetilde{V}_t, \quad t = q,\dots,n-1. \tag{1.1}
\end{equation}
Then they compute the kernel estimator of the LRV of $\{\widetilde{V}_t\}$ because $\widetilde{V}_t$ behaves more like white noise and is less autocorrelated than $\{\hat{V}_t\}$.
\begin{equation}
    \hat{S}_{\widetilde{V}} = \frac{n}{n-d-1}\sum_{r=-n+1}^{n-1}k\left(\frac{r}{m}\right)\hat{\Gamma}_{\widetilde{V}}(r), \tag{1.2}
\end{equation}
where $k(\cdot)$ is the kernel, a real-valued weight function satisfying certain regularity conditions, and $\hat{\Gamma}_{\widetilde{V}}(r) = \frac{1}{n}\sum_{t=r}^{n-1}\widetilde{V}_t\widetilde{V}_{t-r}'$ for $0 \leq r \leq n-1$, $\hat{\Gamma}_{\widetilde{V}}(r) = \hat{\Gamma}_{\widetilde{V}}'(-r)$ for $-n+1 \leq r < 0$ is the sample autocovariance of $\widetilde{V}_t$ at lag $r$. Finally, the LRV estimator of $\{\hat{V}_t\}$ is recolored from $\hat{S}_{\widetilde{V}}$ by inverting the VAR filter:
\begin{equation}
    \hat{S}_{\hat{V}} = \Phi\hat{S}_{\widetilde{V}}\Phi', \quad \Phi = \left(I - \sum_{i=1}^q\hat{A}_i\right)^{-1}. \tag{1.3}
\end{equation}
To maintain numerical stability when calculating the inverse, they suggest adjusting the eigenvalues of $\{\hat{A}_i\}_{i=1}^q$ so that $I - \sum_{i=1}^q\hat{A}_i$ is not too close to singularity.

Based on these three steps, five\footnote{Another important question is which kernel to use. Popular kernels such as the Bartlett, Parzen, and Quadratic Spectral kernel typically underestimate $S_{\hat{V}}$. Vats and Flegal (2022) and Liu and Chan (2025) addressed this issue by proposing kernels with positive bias. Therefore, we omit this question in our paper.} key questions arise:

1). Can we treat the mean of the regressors as zero without loss of generality?

2). How to select the order of the prewhitening VAR?

3). How to estimate the prewhitening VAR?

4). How to select the bandwidth for smoothing?

5). How to perform eigen adjustment if it is needed in the first place?

To provide a well-grounded HAC estimator with satisfactory performance on various metrics, a holistic view of these issues is essential. We believe that questions 1), 3), and 5) are inherently intertwined, as are questions 2) and 4). Existing literature addresses some of these questions while neglecting others, but solutions to one question may introduce new problems in another. Newey and West (1994)\footnote{They claimed their results are invariant to the eigen adjustment rule of Andrews and Monahan (1992) after experimenting it on regressors with mean close to zero. Thus, their recommended HAC estimator omits such an adjustment.} recommended and Den Haan and Levin (2000) used OLS without eigen adjustment to estimate the prewhitening VAR. It is well-known that the OLS estimator of a stationary VAR is not guaranteed to be stationary, not to mention numerical stability. Simulations reveal that using OLS without eigen adjustment can lead to occurrence of large outliers that greatly inflate the bias and variance of the HAC estimator. Andrews and Monahan (1992) used OLS with their eigen adjustment rule to estimate the prewhitening VAR, followed by Lee and Phillips (1994) and Sul, Phillips, and Choi (2005). However, they focused exclusively on zero-mean regressors in their simulations. In practice, it is rare that economic time series have zero mean (e.g., unemployment rate) or that empirical researchers first demean their regressors (e.g., Fama and French 1993). As we will show, the eigen adjustment rule of Andrews and Monahan (1992) can be triggered unnecessarily and harmfully when regressors have nonzero mean. In terms of questions 2) and 4), the order of the prewhiteneng VAR and the bandwidth are two parameters that must be determined to compute the prewhitened kernel estimator. Extensive literature has proposed different ways of selecting the bandwidth (e.g., Andrews 1991, Newey and West 1994, Liu and Chan 2025) but not the order, which is usually predetermined to be one for parsimony. The need for order selection is documented in Lee and Phillips (1994) and Den Haan and Levin (2000), but order selection methods such as AIC and BIC are not applicable to bandwidth selection. To the best of our knowledge, no proposed HAC method can simultaneously select the order and bandwidth besides Xu and Hurvich (2023), whose method is unfortunately limited to the location model $y_t = \mu + u_t$.

The goal of the paper is to propose a new HAC estimator based on the prewhitened kernel estimator and a localized version of the leave-one-out frequency domain cross-validation. We make two main contributions. First, we analyze the role of the regressors's mean in the structure of $V_t$ and in the prewhitening stage. It follows from these results that we cannot treat the mean of the regressors as zero without loss of generality. More importantly, we are the first to study how the mean of the regressors can interfere with the eigen adjustment rule of Andrews and Monahan (1992) and point out certain flaws when regressors have nonzero mean. We recommend using the Burg method (see Brockwell, Dahlhaus, and Trindade 2005) without eigen adjustment to estimate the prewhitening VAR. Second, we propose a localized leave-one-out frequency domain cross-validation suitable to the HAC problem. It not only fills the gap that no cross-validation method has been proposed for bandwidth selection in a general regression setting, but also allows for order selection. Simulations reveal that our method has a satisfactory performance across all metrics: bias, mean-squared error (MSE), and the coverage rate of the confidence interval.

The remainder of the paper is organized as follows. Section II introduces the background of the HAC problem and some assumptions, properties, and concepts that will be used later. Section III reveals some flaws of the eigen adjustment rule of Andrews and Monahan (1992) when regressors have nonzero mean and offers a solution. Section IV introduces a localized leave-one-out frequency domain cross-validation that simultaneously selects the order and bandwidth for the prewhitened kernel estimator and explains the advantages of our method. Section V illustrates the flaws of eigen adjustment and the reliability of our method through Monte Carlo simulations. Section VI strengthens our findings through empirical evidence. Section VII concludes and provides directions for future research.

\section{Background Settings}
In this paper, we consider the general linear regression model with time series regressors. Our goal is to estimate $n\text{Var}(\hat{\beta})$.
\begin{align}
    &Y_t = X_t'\beta + u_t, \quad X_t = (1, x_{1t},\dots,x_{dt})', \quad t=0,\dots,n-1, \notag \\
    &\hat{\beta} = \left(\sum_{t=0}^{n-1}X_tX_t'\right)^{-1}\sum_{t=0}^{n-1}X_tY_t, \quad \text{and} \notag \\
    &n\text{Var}(\hat{\beta}) = E\left(S_n^{-1}\Lambda_nS_n^{-1}\right), \quad S_n = \frac{1}{n}\sum_{t=0}^{n-1}X_tX_t', \quad \Lambda_n = \frac{1}{n}\sum_{i=0}^{n-1}\sum_{j=0}^{n-1}u_iX_iu_jX_j'. \tag{2.1}
\end{align}

The HAC literature has made use of the fact that $\hat{\beta}$ is asymptotically normal:
\begin{equation}
    [E(S_n^{-1})E(\Lambda_n)E(S_n^{-1})]^{-1/2}\sqrt{n}(\hat{\beta}-\beta) \overset{d}{\rightarrow} N(0, I_{d+1}) \quad \text{as} \quad n \to \infty. \tag{2.2}
\end{equation}
Since it is easy to construct a consistent estimator of $\lim_{n\to\infty}E(S_n^{-1})$, usually $S_n^{-1}$ suffices, the goal becomes to estimate
\begin{equation*}
    \lim_{n\to\infty}E(\Lambda_n) = \sum_{r=-\infty}^\infty \Gamma_V(r), \quad V_t = u_tX_t,
\end{equation*}
which is the LRV of $\{V_t\}$. Given $\hat{S}_{\hat{V}}$, an estimator of the LRV of $\{\hat{V}_t = \hat{u}_tX_t\}$, $\hat{u}_t = Y_t-X_t'\hat{\beta}$, the final HAC estimator is expressed as
\begin{equation*}
    n\widehat{\text{Var}}(\hat{\beta}) = \left(\frac{1}{n}\sum_{t=0}^{n-1}X_tX_t'\right)^{-1} \hat{S}_{\hat{V}} \left(\frac{1}{n}\sum_{t=0}^{n-1}X_tX_t'\right)^{-1}.
\end{equation*}

We make the following assumptions on the regressors and error.

\textbf{Assumption}: Let $X_t = (1, x_{1t}, \dots, x_{dt})' \in \mathbb{R}^{d+1}$, $u_t \in \mathbb{R}$, and $E(u_t)=0$. $\{X_t\}_{t=0}^{n-1}$ and $\{u_t\}_{t=0}^{n-1}$ are independent weakly stationary series with absolutely summable autocovariance functions and finite fourth moments, i.e., $E||X_t||_2^4 < \infty$ and $E(u_t^4) < \infty$.

It is helpful to first understand how the structure of $V_t$ is determined by the structure of $X_t$ and $u_t$. Since LRV estimators are constructed on the sample analog $\hat{V}_t$, we then extend our discussion to the relationship between $V_t$ and $\hat{V}_t$.

\textbf{Theorem 2.1}: If $(x_{it}, x_{jt-r}, u_t, u_{t-r})$ is a multivariate normal random vector for $i,j = 1,\dots,d$ and $t,t-r = 0,\dots,n-1$, then $\{V_t\}_{t=0}^{n-1}$ is a zero-mean weakly stationary series. Moreover, the autocovariance of $V_t$ is related to $X_t$ and $u_t$ by
\begin{equation}
    \Gamma_{V}(r) = \gamma_u(r)\Gamma_X(r) + \gamma_u(r)\mu_X\mu_X', \quad \gamma_u(r) = E(u_tu_{t-r}) \ \ \text{and} \ \ \mu_X = E(X_t). \tag{2.3}
\end{equation}

\textbf{Remark}: Theorem 2.1 includes special cases where $x_{it}$, $i=1,\dots,d$, and $u_t$ follow mutually independent weakly stationary ARMA models with normally distributed innovations. Theorem 2.1 applies to most of the simulations in the HAC literature.

\textbf{Theorem 2.2}: If $\sqrt{n}(\hat{\beta}-\beta) = O_p(1)$, then $\hat{V}_t-V_t = O_p(n^{-1/2})$, in particular $\hat{V}_t \overset{p}{\rightarrow} V_t$.

Finally, we introduce some concepts that are used in defining our method in Section IV.

The discrete Fourier transform (DFT) of $\{X_t\}_{t=0}^{n-1}$ is defined as
\begin{equation*}
    J(\omega_j) = \frac{1}{n}\sum_{t=0}^{n-1}X_te^{-i\omega_jt}, \quad j = 0,\dots,n-1,
\end{equation*}
where $J(\omega_j) \in \mathbb{C}^p$ and $\omega_j = 2\pi j/n$ is the $j$-th Fourier frequency. $J(\omega_{n-j}) = \overline{J(\omega_j)}$.

Given $\{J(\omega_j)\}_{j=0}^{n-1}$, its original series $\{X_t\}_{t=0}^{n-1}$ can be recovered by the inverse Fourier transform
\begin{equation*}
    X_t = \sum_{j=0}^{n-1}J(\omega_j)e^{i\omega_jt}, \quad t = 0,\dots,n-1.
\end{equation*}

The periodogram of $\{X_t\}_{t=0}^{n-1}$ is defined as
\begin{equation*}
    I(\omega) = \frac{n}{2\pi}J(\omega)J(\omega)^*,
\end{equation*}
where $*$ denotes the conjugate transpose. ${I(\omega_j)} = I(\omega_{n-j})$.

\section{Eigen Adjustment: Flaws and an Alternative}
In this section, we demonstrate how the eigen adjustment rule of Andrews and Monahan (1992) can be triggered and why it becomes problematic for regressors with nonzero mean. We assume $x_{it}$, $i=1,\dots,d$, $d\geq 1$, and $u_t$ follow mutually independent (weakly) stationary AR($p$) models:
\begin{align*}
    &X_t = (1, x_{1t}, \dots, x_{dt})', \quad x_{it} = \alpha + \sum_{k=1}^p\phi_kx_{t-k} + \varepsilon_{it}, \quad E(x_{it}) = \frac{\alpha}{1-\phi_1-\cdots-\phi_p}, \\
    u_t &= \sum_{k=1}^p\phi_ku_{t-k} + \tilde{\varepsilon}_t, \quad V_t = u_tX_t, \quad (\varepsilon_{1t}, \dots, \varepsilon_{dt}, \tilde{\varepsilon}_t) \overset{i.i.d.}{\sim} N(0,I_{d+1}), \quad t = 0,\dots,n-1.
\end{align*}
$\alpha$ is one of the parameters that controls the mean of $x_{it}$. For algebraic simplicity, the set of AR coefficients is the same for $x_{it}$ and $u_t$. Our setup is a benchmark case of the simulations in existing HAC literature. Previous studies that discuss eigen adjustment mostly set $\alpha = 0$ in their simulations, see Andrews and Monahan (1992) and Sul, Phillips, and Choi (2005). Newey and West (1994) set $\alpha \approx 0.006$, $E(x_t) \approx 0.009$ in experiment A1 and $\alpha \approx 0.001$, $E(x_t) \approx 0.003$ in experiment A2, values that are effectively no different from zero, and they find their results invariant to the eigen adjustment rule of Andrews and Monahan (1992). These suggest that the importance of $\alpha$ has been overlooked.

\subsection{Flaws of Eigen Adjustment}
Eigen adjustment is employed during the prewhitening stage, see equations (1.1) to (1.3) for how prewhitening works. Prewhitening involves estimating a VAR(1)\footnote{Currently, eigen adjustment has only been proposed for the OLS estimator of the prewhitening VAR(1). Hence, our discussion in Section III is restricted to prewhitening using a VAR(1).} on $\{\hat{V}_t\}$ by OLS. Andrews and Monahan (1992) suggest using an eigenvalue adjusted version of the OLS estimator of the prewhitening VAR(1) so that it is numerically stable to invert the VAR(1) filter in the final step of recoloring. We begin by describing how prewhitening works on $V_t$.

\textbf{Theorem 3.1}: When $p=1$, $V_t$ is a stationary VAR(1) model of the form
\begin{equation*}
    V_t = AV_{t-1} + \eta_t, \quad A = \begin{bmatrix}
        \phi & 0_{1 \times d} \\
        \alpha\phi1_{d \times 1} & \phi^2I_d
    \end{bmatrix} \quad \eta_t = \begin{bmatrix}
        \tilde{\varepsilon}_t \\
        (\varepsilon_{1t}\phi u_{t-1} + \tilde{\varepsilon}_tx_{1t}, \dots, \varepsilon_{dt}\phi u_{t-1} + \tilde{\varepsilon}_tx_{dt})'
    \end{bmatrix}.
\end{equation*}
The OLS estimator $\hat{A}$ of $A$ is consistent. When $p \geq 2$, $V_t$ is not a VAR model of any order.

\textbf{Remark}: $V_t$ can be fully prewhitened by a VAR(1) for $p=1$, but not for $p \geq 2$.

As the eigen adjustment rule of Andrews and Monahan (1992) involves singular values, we also analyze how the eigenvalues and singular values of $A$ depend on $\alpha$, in other words the mean of the regressors.

\textbf{Lemma 3.1}: The eigenvalues of $A$ are $\phi$ with multiplicity 1 and $\phi^2$ with multiplicity $d$. The singular values of $A$ are $\phi^2$, $|s_1|$, $|s_2|$ with multiplicity $d-1$, 1, 1 respectively, where
\begin{equation*}
    s_1, s_2 = \frac{\phi}{2}\sqrt{\alpha^2d+(\phi+1)^2} \ \pm \ \frac{\phi}{2}\sqrt{\alpha^2d+(\phi-1)^2}.
\end{equation*}

\textbf{Remark}: $d$ is the number of nonconstant regressor and we assume $d \geq 1$. It is important to recognize that the eigenvalues of $A$ are independent of $\alpha$ whereas two singular values of $A$, $|s_1|$ and $|s_2|$, depend on $\alpha$ and always exist for $d \geq 1$. Therefore, we set $d=1$ without loss of generality for the remainder of the section. In this case the singular value $\phi^2$ vanishes.

Next, we demonstrate the eigen adjustment rule of Andrews and Monahan (1992) on the coefficient matrix $A$ derived in Theorem 3.1. It deviates from $\hat{A}_{\hat{V}}$, the OLS estimator of the prewhitening VAR(1) that is used in practice, in three ways. First, $A$ is the true coefficient matrix instead of the OLS estimator $\hat{A}$. Second, $A$ is derived under $p=1$. By using $A$ we implicitly assume $p=1$ and will omit the discussion for $p \geq 2$. Third, $A$ is constructed from $V_t$ rather than $\hat{V}_t$. The first two deviations are justified by Theorem 3.1. On the one hand, since $\hat{A}$ is consistent, $A$ is representative of how eigen adjustment on $\hat A$ will behave when the sample size is large. On the other hand, since $V_t$ is not a VAR(1) model for $p \geq 2$, the OLS estimator $\hat{A}$ has no closed form. Therefore, demonstrating the eigen adjustment rule on $\hat{A}$ will not provide any meaningful intuition, but our conclusions still apply, as shown in section V for $p=2,3$. The third deviation is justified by $\hat{V}_t \overset{p}{\rightarrow} V_t$ established in Theorem 2.2. Furthermore, we validate our results derived under $V_t$ and $A$ using $\hat{V}_t$ and $\hat{A}_{\hat{V}}$ in simulations.

The eigen adjustment rule of Andrews and Monahan (1992, p957) is to conduct a singular value decomposition on $A$ (actually, on $\hat{A}_{\hat{V}}$) and bound the eigenvalues $\{\lambda_i(A)\}$ by restricting its singular values $\{\sigma_i(A)\}$ through the relation
\begin{equation*}
    \min_i\sigma_i(A) \leq \min_i|\lambda_i(A)| \leq \max_{i}|\lambda_i(A)| \leq \max_i\sigma_i(A).
\end{equation*}

More specifically, let B and C be two orthogonal matrices whose columns are eigenvectors of $AA'$ and $A'A$. $\Delta = B'AC$ is a diagonal matrix with $s_1$ and $s_2$ on the diagonal. If $\alpha \neq 0$,
\begin{equation*}
    A = \begin{bmatrix}
        \phi & 0 \\
        \alpha\phi & \phi^2
    \end{bmatrix} \ \ B = \begin{bmatrix}
        \dfrac{\alpha\phi^2}{B_1} & \dfrac{\alpha\phi^2}{B_2} \\
        \dfrac{s_1^2-\phi^2}{B_1} & \dfrac{s_2^2-\phi^2}{B_2}
    \end{bmatrix} \ \ C = \begin{bmatrix}
        \dfrac{s_1^2-\phi^4}{C_1} & \dfrac{s_2^2-\phi^4}{C_2} \\
        \dfrac{\alpha\phi^3}{C_1} & \dfrac{\alpha\phi^3}{C_2}
    \end{bmatrix} \ \ \Delta_{\alpha \neq 0} = \begin{bmatrix}
        s_1 & 0 \\
        0 & s_2
    \end{bmatrix}.
\end{equation*}
$B_i = \sqrt{\alpha^2\phi^4+(s_i^2-\phi^2)^2}$, $C_i = \sqrt{\alpha^2\phi^6+(s_i^2-\phi^4)^2}$, $i=1,2$, are the normalizing constants. If $\alpha = 0$,
\begin{equation*}
    A = \begin{bmatrix}
        \phi & 0 \\
        0 & \phi^2
    \end{bmatrix} \quad B = C = \begin{bmatrix}
        1 & 0 \\
        0 & 1
    \end{bmatrix} \quad \Delta_{\alpha = 0} = \begin{bmatrix}
        \phi & 0 \\
        0 & \phi^2
    \end{bmatrix}.
\end{equation*}
To restrict the singular values of $A$, a matrix $\tilde{\Delta}$ is constructed by replacing the diagonal elements of $\Delta$ that are larger than 0.97 with 0.97 and those smaller than $-0.97$ with $-0.97$. The eigenvalue adjusted version is $A^{\text{adj}} = B\tilde{\Delta}C'$, which guarantees the eigenvalues of $A^{\text{adj}}$ to be less than 0.97 in magnitude so that $I-A^{\text{adj}}$ is not too close to singularity.

It follows that their procedure is triggered when the singular values of $A$ exceed 0.97 and it should be triggered only if the eigenvalues of $A$ exceed 0.97 in magnitude in the first place. However, Table 1 indicates that when $\alpha \neq 0$, eigen adjustment is triggered even though the eigenvalues of $A$ have magnitude within $[-0.97, 0.97]$, which is contrary to the original goal.

\begin{center}
    [Table 1]
\end{center}

When $\phi=0.5$, although the largest eigenvalue of $A$ is 0.47 below the threshold 0.97, the largest singular value already exceeds 0.97, which unnecessarily triggers the procedure. Moreover, eigen adjustment is harmful in the sense that it distorts $A$ by as much as 56.4\%.

The flaw lies in the fact that the singular values of $A$, $|s_1|$ and $|s_2|$, depend on $\alpha$ whereas the eigenvalues of $A$, $\phi$ and $\phi^2$, are independent of $\alpha$. In fact, the largest singular value $|s_1|$ increases with $\alpha$. Since singular values are the triggers of the procedure and it is possible for the eigenvalues to be small but singular values to be large with a small $\phi$ and a large $\alpha$, eigen adjustment can be triggered unnecessarily and harmfully.

Finally, we address the limitation that our discussion is based on $V_t=u_tX_t$ and $A$, neither of which is available in practice, by validating our findings in Table 2. Table 2 reflects what happens in practice if the eigen adjustment rule of Andrews and Monahan (1992) is applied to $\hat{A}_{\hat{V}}$, the OLS estimator of the prewhitening VAR(1) constructed using $\hat{V}_t=\hat{u}_tX_t$, where $\hat{u}_t$ denotes the estimated error from regressing $Y_t$ on $X_t$ under $Y_t = X_t'\beta+u_t$ and $\beta=0$. This is also the setting considered in Andrews and Monahan (1992).

\textbf{Remark}: From equation (2.1), the distribution of $n\text{Var}(\hat{\beta})$ is invariant to $\beta$, so we set $\beta = 0$ in our simulations without loss of generality for the remainder of the paper.

\begin{center}
    [Table 2]
\end{center}

Three major observations confirm our findings. First, consistent with Lemma 3.1, the eigenvalues of $\hat{A}_{\hat{V}}$ are close to $\phi$ and $\phi^2$, the singular values of $\hat{A}_{\hat{V}}$ are close to $|s_1|$ and $|s_2|$. Second, eigen adjustment on $\hat{A}_{\hat{V}}$ is often unnecessarily triggered. When $\phi=0.9$, it happens almost every time because the singular values of $\hat{A}_{\hat{V}}$ are inflated by $\alpha$ to exceed 0.97 while the eigenvalues of $\hat{A}_{\hat{V}}$ remain below the threshold on average. Third, eigen adjustment still distorts $\hat{A}_{\hat{V}}$ by as much as 60.7\%. As the prewhitened kernel estimator relies heavily on $\hat{A}_{\hat{V}}$, frequent adjustment with sizable distortion can greatly inflate the variance of the estimator, as we will show in Example 1, Figures 3 and 4.

In summary, we have enough evidence to believe that in general:

1). The singular values of $\hat{A}_{\hat{V}}$ depend on the mean of the regressors.

2). It is possible for the eigenvalues of $\hat{A}_{\hat{V}}$ to be small but the singular values to be large.
\indent\phantom{2).} Hence, eigen adjustment can be triggered unnecessarily and is potentially harmful.

\subsection{An Alternative to Eigen Adjustment}
The need for eigen adjustment stems from two facts. Theoretically, since $x_{it}$, $i=1,\dots,d$, and $u_t$ follow mutually independent stationary AR models, $V_t$ is stationary by Theorem 2.1, and the sample analog $\hat{V}_t$ is asymptotically equivalent to a stationary series by Theorem 2.2. Generally speaking, most HAC estimators are theoretically valid only under stationary $\hat{V}_t$. Consequently, $\hat{A}_{\hat{V}}$ should correspond to a stationary VAR(1) model, that is the eigenvalues of $\hat{A}_{\hat{V}}$ should be less than one in magnitude (we sometimes omit the phrase ``in magnitude"). However, estimating a VAR model on a stationary series by OLS does not necessarily yield coefficient matrices that correspond to a stationary VAR model. Practically, if the eigenvalues of $\hat{A}_{\hat{V}}$ are close to one, the eigenvalues of $I-\hat{A}_{\hat{V}}$ are close to zero, $I-\hat{A}_{\hat{V}}$ is nearly singular and calculating its inverse is numerically unstable, i.e., some entries may blow up. Example 1 demonstrates how the prewhitened kernel HAC estimator using OLS without eigen adjustment to estimate the prewhitening VAR(1) can go wrong.

\textbf{Example 1}: $Y_t=X_t'\beta+u_t$, $X_t=(1,x_{t})$, $x_{t}=1+0.95x_{t-1}+\varepsilon_t$, $u_t=0.95u_{t-1}+\tilde{\varepsilon}_t$, $(\varepsilon_t,\tilde{\varepsilon}_t) \sim N(0,I_2)$, $n=200$, and 1000 repetitions. The estimand of interest is $n\text{Var}(\hat{\beta})_{22}$.

\begin{center}
    [Figure 1] \hspace{15em} [Figure 2]
\end{center}

Figures 1 and 2 indicate that large outliers occur when the eigenvalues of $\hat{A}_{\hat{V}}$ are close to one. For example, in repetition 343, the largest eigenvalue of $\hat{A}_{\hat{V}}$ is 1.0003, which violates the stationary condition. Furthermore, $I-\hat{A}_{\hat{V}}$ is almost singular and calculating its inverse lead to
\begin{equation*}
    \hat{A}_{\hat{V}} = \begin{bmatrix}
        0.912 & 0.002 \\
        -1.45 & 1.03
    \end{bmatrix} \quad \text{and} \quad (I-\hat{A}_{\hat{V}})^{-1} = \begin{bmatrix}
          1940 & -117 \\
          88359 & -5344
    \end{bmatrix}.
\end{equation*}
Consequently, the HAC estimator blows up. The variance of $n\widehat{\text{Var}}(\hat{\beta})_{22}$ is $4.2\times10^{10}$.
\begin{align*}
    n\widehat{\text{Var}}(\hat{\beta}) &= \left(\frac{1}{n}\sum_{t=0}^{n-1}X_tX_t'\right)^{-1} \Phi \hat{S}_{\tilde{V}} \Phi' \left(\frac{1}{n}\sum_{t=0}^{n-1}X_tX_t'\right)^{-1}, \quad \Phi = (I-\hat{A}_{\hat{V}})^{-1} \\
    &= \begin{bmatrix}
        2.7\times10^9 & -1.3\times10^8 \\
        -1.3\times10^8 & 6.4\times10^6
    \end{bmatrix}.
\end{align*}

Although adjusting the eigenvalues of $\hat{A}_{\hat{V}}$ to fall within $[-0.97, 0.97]$ satisfies both needs, the only eigen adjustment rule proposed to date, by Andrews and Monahan (1992), has been found to be problematic for regressors with nonzero mean. In Example 1,

\begin{center}
    [Figure 3] \hspace{15em} [Figure 4]
\end{center}
compared with OLS without eigen adjustment that produces one large outlier, OLS with the eigen adjustment rule of Andrews and Monahan (1992) produces numerous moderate outliers that inflates the variance of $n\widehat{\text{Var}}(\hat{\beta})_{22}$ to $4.5\times10^6$. This is precisely because eigen adjustment is triggered too frequently and each time it introduces sizable distortion to $\hat{A}_{\hat{V}}$.

There are two possible ways to address this issue. One either proposes a more appropriate eigen adjustment rule or chooses a method to estimate the prewhitening VAR so that eigen adjustment can be avoided. The former approach suffers the difficulty that when higher order prewhitening VAR($p$) is used, the eigenvalues of $p$ coefficient matrices need to be adjusted. Therefore, we adopt the latter approach and use the Burg method without eigen adjustment to estimate the prewhitening VAR. As the Burg method for vector autoregression is rather involved but well established, we do not provide details here. Instead, we cite existing results in support of our choice and demonstrate its reliability through Example 1.

\textbf{Theorem 3.2}: The Burg estimators $\{\hat{B}_i\}_{i=1}^p$ of a stationary VAR($p$) model is guaranteed to be stationary, that is the eigenvalues of $B$ are less than one in magnitude, where
\begin{equation*}
    B = \begin{bmatrix}
        \hat{B}_1 \quad \cdots\cdots \quad \hat{B}_p \\
        I_{(d+1)(p-1)} \quad 0_{(d+1)(p-1)\times(d+1)}
    \end{bmatrix}.
\end{equation*}

\textbf{Remark}: The Burg method was originally proposed by Burg (1975) for spectral density estimation of univariate stationary time series. Brockwell, Dahlhaus, and Trindade (2005) generalized the Burg method to multivariate stationary time series. A proof of Theorem 3.2 is given by Morf, Vieira, and Kailath (1978).

\begin{center}
    [Figure 5] \hspace{15em} [Figure 6]
\end{center}

In Example 1, a small outlier occurs and the largest eigenvalue of $\hat{B}_1$, the Burg estimator of the prewhitening VAR(1) without eigen adjustment, is only 0.003 above the threshold 0.97. The variance of $n\widehat{\text{Var}}(\hat{\beta})_{22}$ is reduced to 218.

\section{Automatic Order and Bandwidth Selection}
In this section, we introduce a localized leave-one-out frequency domain cross-validation (FDCV) that selects the order $q$ and the bandwidth $m$ for the prewhitened kernel estimator. FDCV is first and mainly used for bandwidth selection in spectral density estimation because it focuses on estimating the spectral density function over a set of frequencies, not at a single frequency zero that the HAC problem concerns. We begin by giving an overview of FDCV and discussing how it can be adapted to the HAC problem while enabling order selection.

\subsection{An Overview of Frequency Domain Cross-Validation}
Beltrao and Bloomfield (1987) first proposed the univariate cross-validated log likelihood (CVLL) function
\begin{equation*}
    \sum_{j=1}^{\lfloor n/2 \rfloor}\left[\log\hat{f}_{-j}(\omega_j, m) + \frac{I(\omega_j)}{\hat{f}_{-j}(\omega_j, m)}\right]
\end{equation*}
for the leave-one-out discrete periodogram average estimator, where $\hat{f}_{-j}(\omega_j, m)$ is a weighted sum of $I(\omega_{j-k})$, $k = \pm 1, \dots, \pm m$, excluding $I(\omega_j)$. Beltrao and Bloomfield (1987) proved that their CVLL is asymptotically equivalent to the mean integrated squared error (MISE), which measures the quality of the spectral density (spectrum) estimator over a set of frequencies. Since the optimal bandwidth $m^*$ is the minimizer of CVLL chosen from a search region of $m$, it emphasizes the quality of the spectrum estimator over a set of frequencies as well.

Adapting cross-validation to the HAC problem requires three generalizations. First, a multivariate analog of CVLL is needed to accommodate multivariate spectrum estimators. Recall that the goal is to estimate $S_V=2\pi f_V(0)$, where $V_t=u_tX_t$. If $\hat{f}_{-j}$ is a scalar, then HAC estimation is limited to $X_t=1$, in other words regressions of the type $y_t = \mu + u_t$. Second, cross-validation should select the bandwidth that prioritizes the goodness of fit of the spectrum estimator at zero, or at least over a set of frequencies around zero. The univariate CVLL of Beltrao and Bloomfield (1987) is defined over the set $\{\omega_1,\dots,\omega_{\lfloor n/2 \rfloor}\}$, $\omega_j = 2\pi j/n$, which is not local to zero. Third, even if the bandwidth prioritizes the goodness of fit of the spectrum estimator at zero, estimators such as the discrete periodogram average have been found to perform poorly for strongly autocorrelated $V_t$. Specifically, we want cross-validation to include the prewhitened kernel estimator because it yields the best performance in general. This, however, introduces a new problem that the order of the prewhitening VAR needs to be selected in conjunction with the bandwidth.

Our generalization builds upon the following literature. First, Hurvich (1985) extended the leave-one-out definition of Beltrao and Bloomfield's (1987) discrete periodogram average to any spectrum estimator. Cross-validation is no longer used to choose a single parameter, but the estimator as a whole. Since prewhitened kernel estimators with different orders and bandwidths can be regarded as distinct estimators, by using the generalized leave-one-out definition of Hurvich (1985), it not only allows cross-validation to include the prewhitened kernel estimator, but also to simultaneously select its order and bandwidth. Next, we adopt a multivariate analog of CVLL suggested by Robinson (1991, p1346):
\begin{equation*}
    \sum_{j=1}^n\left\{\log\det\hat{f}_{-j}(\omega_j) + \text{tr}\left[I(\omega_j)\hat{f}_{-j}^{-1}(\omega_j)\right]\right\}.
\end{equation*}
Yet Robinson (1991) only provided theories of the univariate CVLL for the leave-two-out discrete periodogram average estimator and the multivariate CVLL is still defined over a set of frequencies not local to zero. Finally, we borrow the idea of using the first $\lfloor(n/2)^c\rfloor$ frequencies in the cross-validation from Xu and Hurvich (2023), who proposed a localized cross-validated function
\begin{equation*}
    \sum_{j=1}^{\lfloor(n/2)^c\rfloor} \left\{\left[\log\hat{f}_{-j}(\omega_j) - \log I(\omega_j) - C\right]^2 - \frac{\pi^2}{6}\right\}
\end{equation*}
by adding an exponent $c \in (0,1)$. Asymptotically, only Fourier frequencies around zero are considered as $\lim_{n\to\infty}\{2\pi\lfloor(n/2)^c\rfloor\}/n = 0$.

We now formally introduce our FDCV method for any leave-one-out spectrum estimator suitable to the HAC problem in a general linear regression setting.

\subsection{A Localized Frequency Domain Cross-Validation for HAC \\ Estimation}
Given $\{\hat{V}_t\}_{t=0}^{n-1}$, we apply DFT to $\{\hat{V}_t\}_{t=0}^{n-1}$ and obtain $\{J(\omega_k)\}_{k=0}^{n-1}$, where
\begin{equation*}
    J(\omega_k) = \frac{1}{n}\sum_{t=0}^{n-1}\hat{V}_te^{-i\omega_kt}, \quad k = 0, \dots, n-1.
\end{equation*}
To leave $j$ out, let $J(\omega_n) = J(\omega_0)$, we define $\{J^{-j}(\omega_k)\}_{k=0}^{n-1}$, $j = 1,\dots,n-1$, as
\begin{equation}
    J^{-j}(\omega_k) = \begin{cases}
        J(\omega_k), &k \neq j \ \text{and} \ k \neq n-j \\
        \frac{1}{2}[J(\omega_{j-1}) + J(\omega_{j+1})], &k=j \ \text{and} \ k=n-j
    \end{cases}. \tag{4.1}
\end{equation}
The leave-$j$-out dataset $\{\hat{V}_t^{-j}\}_{t=0}^{n-1}$ is recovered from $\{J^{-j}(\omega_k)\}_{k=0}^{n-1}$ by the inverse Fourier transform
\begin{equation*}
    \hat{V}_t^{-j} = \sum_{k=0}^{n-1}J^{-j}(\omega_k)e^{i\omega_kt}, \quad t = 0, \dots, n-1.
\end{equation*}
Any spectrum estimator based on $\{\hat{V}_t^{-j}\}_{t=0}^{n-1}$ is the leave-one-out spectrum estimator $\hat{f}_{-j}$.

If $\{\hat{V}_t\}_{t=0}^{n-1}$ has nonzero mean, then the above procedure is not invariant under adding or subtracting a constant because
\begin{equation*}
    J(\omega_0) = \frac{1}{n}\sum_{t=0}^{n-1}\hat{V}_t \quad \text{and} \quad J(\omega_1) = \frac{1}{2}[J(\omega_0) + J(\omega_2)].
\end{equation*}
In case researchers need to demean $\{\hat{V}_t\}_{t=0}^{n-1}$, $\{J^{-j}(\omega_k)\}_{k=0}^{n-1}$ can be redefined as
\begin{equation}
    J^{-j}(\omega_k) = \begin{cases}
        J(\omega_k), &k \neq j \ \text{and} \ k \neq n-j \\
        \frac{1}{2}[J(\omega_{j-1}) + J(\omega_{j+1})], & k = j \ \text{and} \ n-j, \ j \notin \{1, n-1\} \\
        J(\omega_2), &k = 1, \ j \in \{1, n-1\} \\
        J(\omega_{n-2}), &k = n-1, \ j \in \{1, n-1\}
    \end{cases}. \tag{4.2}
\end{equation}

\textbf{Remark}: It follows from the first order condition of the OLS estimator $\hat{\beta}$ that the sample mean of $\hat{V}_t$ is zero. Hence, (4.2) is unnecessary and we use (4.1). Casini and Perron (2024) discussed HAC estimation under $E(\hat{V}_t) \neq 0$, e.g., when the regression model is misspecified or $\beta$ is time-varying, in which case demeaning is potentially preferred.

We consider a class of VAR prewhitened Parzen kernel estimators
\begin{align*}
    \hat{f}_{\hat{V}}(\omega_j, q, m) = \frac{1}{2\pi}\Phi(e^{-i\omega_j}) \left[\frac{n}{n-d-1}\sum_{r=-n+1}^{n-1}k\left(\frac{r}{m}\right)\hat{\Gamma}_{\widetilde{V}}(r)e^{-i\omega_jr}\right] \Phi^*(e^{-i\omega_j}),& \quad \text{where} \\
    \Phi(z) = \left(I - \sum_{k=1}^q\hat{A}_kz^k\right)^{-1}, \quad k(x) = \begin{cases}
        1-6x^2+6|x|^3, &0 \leq |x| \leq 0.5 \\
        2(1-|x|)^3, &0.5 \leq |x| \leq 1 \\
        0, &\text{otherwise}
    \end{cases},& \\
    \widetilde{V}_t = \hat{V}_t - \sum_{k=1}^q\hat{A}_k\hat{V}_{t-k}, \quad q \in \{1,2\}, \ \ \text{and} \quad m \in \{1,\dots,\lfloor4(n/100)^{\frac{2}{9}}\rfloor\}.&
\end{align*}
$\{\hat{A}_k\}_{k=1}^q$ are the Burg estimators of the prewhitening VAR($q$) fitted to $\hat{V}_t$. $\hat{f}_{-j}(\omega_j, q, m)$ is obtained by replacing $\hat{V}_t$ in the calculations of $\{\hat{A}_k\}_{k=1}^q$ and $\widetilde{V}_t$ with $\hat{V}_t^{-j}$.

For each $(q,m)$, $\hat{f}_{-j}(\omega_j, q, m)$ is substituted into a localized multivariate CVLL
\begin{equation*}
    \text{CVLL}_c = \sum_{j=1}^{\lfloor (n/2)^c \rfloor} \left\{\log\det\hat{f}_{-j}(\omega_j, q, m) + \text{tr}\left[I(\omega_j)\hat{f}^{-1}_{-j}(\omega_j, q, m)\right]\right\},
\end{equation*}
where $c \in (0,1)$, $\text{tr}(\cdot)$ denotes the trace of a matrix, and $I(\omega_j)$ is always constructed on $\hat{V}_t$. Whichever $(q^*, m^*)$ that minimizes CVLL$_c$ is selected and the final HAC estimator
\begin{equation*}
    n\widehat{\text{Var}}(\hat{\beta}) = \left(\frac{1}{n}\sum_{t=0}^{n-1}X_tX_t'\right)^{-1}2\pi\hat{f}_{\hat{V}}(0, q^*, m^*)\left(\frac{1}{n}\sum_{t=0}^{n-1}X_tX_t'\right)^{-1}.
\end{equation*}

To implement this method, researchers must specify a class of spectrum estimators for CVLL$_c$ to choose from. Our class of VAR prewhitened kernel estimators of order 1 to 2 is illustrative rather than restrictive. In general, the class may include VARMA prewhitened kernel estimators of higher orders, as suggested by Lee and Phillips (1994), or nonparametric estimators such as the discrete periodogram average. This flexibility allows CVLL$_c$ not only to simultaneously select the order and bandwidth, but also to select between parametric and nonparametric estimators. Xu and Hurvich (2023) provided an example of the latter.

\subsection{Discussion on Order and Bandwidth Selection}
For bandwidth selection, the two main types are cross-validation and the plug-in method. Andrews (1991) first proposed a plug-in method to automatically determine the bandwidth and argued that cross-validation is not suitable for the HAC problem because it focuses on estimating the spectral density over an interval instead of at zero. Consequently, the plug-in method has dominated the HAC literature. Here, we fill this gap. The plug-in method first specifies a weight vector $w$, then obtains a proxy for the optimal bandwidth that minimizes the MSE $w'(\hat{S}_{\hat{V}}-S_{\hat{V}})w$, and finally estimates that bandwidth. This method's drawbacks are that it has only been used to determine the bandwidth, since minimizing over both order and bandwidth is rather complicated, and the output is sensitive to $w$. By contrast, our FDCV method can determine the order and bandwidth and only requires choosing the exponent $c$.

The need for order selection is most evident under the location model $Y_t=\mu+u_t$, $u_t \sim \text{ARMA}(p,q)$. In this case, $V_t=u_t$ and order selection should be employed to choose a prewhitening ARMA($p,q$) filter. More empirical evidence on the need for order selection can be found in Lee and Phillips (1994) and Den Haan and Levin (2000).

Asymptotic theory on the order and bandwidth selected by CVLL$_c$ is beyond the scope of this paper. Robinson (1991) provided the most recent results on the bandwidth selected by the univariate CVLL for the discrete periodogram average. His proof relied crucially on Parseval’s formula, which equates the sum of squares of a sequence with the sum of squares of its Fourier transform. This identity requires summation over all values but CVLL$_c$ only sums over the first $\lfloor(n/2)^c\rfloor$ frequencies. Thus, generalizing his results to a localized version is already a nontrivial task, not to mention that CVLL$_c$ is multivariate and accommodates any spectrum estimator. To compensate, in the following two sections we evaluate our method through extensive simulations and three real-world examples.

\section{Monte Carlo Evidence}
In this section, we illustrate the flaws of eigen adjustment and prove the reliability of our method by comparing four HAC standard error estimators in simulations. QS denotes the quadratic spectral (QS) kernel estimator as in Andrews (1991). AM-PW denotes the VAR(1) prewhitened QS kernel estimator as in Andrews and Monahan (1992). The prewhitening VAR(1) is estimated by OLS with eigen adjustment. AM-PW$^\text{unadj}$ is the same as AM-PW except without eigen adjustment. Our method, referred to as CVLL, is the VAR prewhitened Parzen kernel estimator of order 1 to 2 and bandwidth 1 to $\lfloor4(n/100)^{2/9}\rfloor$. The prewhitening VAR is estimated by Burg without eigen adjustment. The order and bandwidth are selected by CVLL$_c$ with $c=0.8$. As shown in Table 9 of Xu and Hurvich (2023), this choice of $c$ balances performance with computational expense.

We consider the linear regression model with an intercept and three regressors:
\begin{equation*}
    y_t = \beta_0 + \sum_{i=1}^3\beta_ix_{it} + u_t, \quad E(u_t) = 0, \quad t=0,\dots,n-1.
\end{equation*}
Our estimand of interest is $n\text{Var}(\hat{\beta}_1)$. Regressors and errors are mutually independent and stationary with sample sizes of 100 and 200. There are 1000 repetitions in each scenario. To measure the performance, we calculate the bias, variance, and MSE of each HAC estimator. The true value $n\text{Var}(\hat{\beta}_1)$ is taken to be $n$ times the variance of the one thousand simulated $\hat{\beta}_1$ across repetitions. We further calculate the confidence interval for $\beta_1$ at nominal rates $100(1-\alpha)\%$, $\alpha = 0.1, 0.05, 0.01$,
\begin{equation*}
    I = [\hat{\beta}_1 - z_{\frac{\alpha}{2}}\hat{\text{SE}}(\hat{\beta}_1), \ \hat{\beta}_1 + z_{\frac{\alpha}{2}}\hat{\text{SE}}(\hat{\beta}_1)], \quad \hat{\text{SE}}(\hat{\beta}_1) = \sqrt{\widehat{\text{Var}}(\hat{\beta}_1)},
\end{equation*}
and examine the width $2z_{\alpha/2}\hat{\text{SE}}(\hat{\beta}_1)$ and the coverage rate $1_{\beta_1 \in I}$ of the confidence interval. $z_{\alpha/2}$ is the two-sided critical value of the standard normal distribution at significance level $\alpha$. Its use follows from equation (2.2).

\subsection{Regressors and Errors are AR(1) and AR(2)}
First, we compare the four HAC estimators under AR(1) regressors and errors. Results with sample size $n=100$ are reported in Tables 3 and 4.
\begin{equation*}
    x_{it} = \alpha + \phi x_{it-1} + \varepsilon_{it}, \quad u_t = \phi u_{t-1} + \tilde{\varepsilon}_t, \quad (\varepsilon_{1t}, \varepsilon_{2t}, \varepsilon_{3t}, \tilde{\varepsilon}_t) \sim N(0,I_4).
\end{equation*}

\begin{center}
    [Table 3] \hspace{15em} [Table 4]
\end{center}

When $\alpha=0$, regressors have zero mean and eigen adjustment is rarely triggered. Under moderate degree of autocorrelation ($\phi=0.3$ and 0.6), the four estimators perform similarly. As regressors and errors become close to nonstationarity ($\phi=0.9$ and 0.95), AM-PW$^\text{unadj}$ produces a few large outliers that greatly inflate the bias and variance. The reported bias of 0.287 does not indicate that AM-PW$^\text{unadj}$ is less biased on average, but that a few large positive bias offset the entire negative bias. Due to order selection, CVLL performs slightly worse (by less than 2\%) in terms of the coverage rate. There is a pattern that as $\phi$ increases, CVLL chooses more prewhitening VAR(1) over VAR(2), and it makes this optimal choice (by Theorem 3.1) around 50\% of the time when $\phi=0.9$ and 0.95.

Under $\alpha=1$, $\phi=0.3$ and 0.6, the four estimators still perform similarly. However, when regressors and errors are close to nonstationarity ($\phi=0.9$ and 0.95), eigen adjustment often introduces sizable distortion that causes AM-PW to produce numerous moderate outliers. Compared with AM-PW under $\alpha=0$, $\phi=0.9$ and 0.95, besides bias and variance inflation, the width of the confidence interval is also inflated. Although this leads to a sharp increase in the coverage rate, such wide intervals are practically meaningless. Large outliers of AM-PW$^\text{unadj}$ and the pattern in the average order selected by CVLL are persistent.

Since AM-PW and AM-PW$^\text{unadj}$ use a fixed prewhitening VAR(1), which is the optimal filter in this example, the benefits of CVLL, particularly its use of Burg and order selection, are more evident for a broader class of data. Next, we compare the four HAC estimators under AR(2) regressors and errors. Results with $n=100$ are reported in Tables 5 and 6.
\begin{equation*}
    x_{it} = \alpha + \sum_{k=1}^2\frac{\phi}{2}x_{it-k} + \varepsilon_{it}, \quad u_t = \sum_{k=1}^2\frac{\phi}{2}u_{t-k} + \tilde{\varepsilon}_t, \quad (\varepsilon_{1t}, \varepsilon_{2t}, \varepsilon_{3t}, \tilde{\varepsilon}_t) \sim N(0,I_4).
\end{equation*}
This AR(2) model is designed to approach nonstationarity as $\phi$ increases.

\begin{center}
    [Table 5] \hspace{15em} [Table 6]
\end{center}

When $\alpha=0$, CVLL generally performs the best with significantly higher coverage rates. This is achieved not by producing outliers that inflate the width of the confidence interval, but by producing the least biased estimators for all values of $\phi$ through order selection. As $\phi$ increases, CVLL chooses more prewhitening VAR(2) over VAR(1), and it makes this choice from 95\% of the time to every time when $\phi=0.9$ and 0.95.

When $\alpha=1$, the flaws of eigen adjustment and the advantages of CVLL remain evident. Abandoning eigen adjustment and only using OLS to estimate the prewhitening VAR is undesired because it leads to substantially lower coverage rates, as shown by AM-PW$^\text{unadj}$. Another important observation is that across all four examples, although CVLL selects from a class of prewhitened kernel estimators, its variance is not heavily inflated (note that part of its variance comes from order selection rather than prewhitening). Consequently, CVLL's MSE is comparable to, and sometimes even lower than that of the kernel estimator AM. Variance inflation of prewhitened kernel estimators has been documented in Andrews and Monahan (1992) and Sul, Phillips, and Choi (2005). Using Burg to estimate the prewhitening VAR seems to mitigate this issue.

Here, CVLL performs well is partly because its search region includes the prewhitening VAR(2) filter. To test the robustness of our method, we finally compare the four estimators in cases where the autocorrelation structure of $\hat{V}_t$ is not well captured by a VAR(2).

\subsection{Regressors and Errors are AR(3) and MA(2)}
The AR(3) and MA(2) data-generating processes (DGP) with $n=200$ are of the form
\begin{align*}
    \text{AR}(3):& \quad x_{it} = \alpha + \sum_{k=1}^3\frac{\phi}{3}x_{it-k} + \varepsilon_{it}, \quad u_t = \sum_{k=1}^3\frac{\phi}{3}u_{t-k} + \tilde{\varepsilon}_t, \quad (\varepsilon_{1t}, \varepsilon_{2t}, \varepsilon_{3t}, \tilde{\varepsilon}_t) \sim N(0,I_4), \\
    \text{MA}(2):& \quad x_{it} = \alpha + \sum_{k=1}^2\phi_k\varepsilon_{it-k} + \varepsilon_{it}, \quad u_t = \sum_{k=1}^2\phi_k\tilde{\varepsilon}_{t-k} + \tilde{\varepsilon}_t, \quad (\varepsilon_{1t}, \varepsilon_{2t}, \varepsilon_{3t}, \tilde{\varepsilon}_t) \sim N(0,I_4).
\end{align*}

\begin{center}
    [Table 7] \hspace{7.5em} [Table 8] \hspace{7.5em} [Table 9]
\end{center}
$\phi_1$ and $\phi_2$ are chosen such that the second-order autocorrelation is much stronger than the first. Since an MA(2) model is invertible to a VAR($\infty$), intuitively prewhitening VAR(3) and VAR($\infty$) filters should be used under AR(3) and MA(2) DGPs respectively. Yet the former is outside the search region and the latter is practically infeasible. Under this circumstance, CVLL almost always selects a prewhitening VAR(2) and our conclusions still apply.

\section{Empirical Evidence}
In this section, we provide three empirical examples to strengthen our findings that eigen adjustment can be triggered unnecessarily and that HAC standard errors, hence statistical significance, are sensitive to the estimation method. The four standard error estimators: QS, AM-PW, AM-PW$^{\text{unadj}}$, and CVLL are described in section V.

\subsection{Unemployment and GDP}
The variable log GDP per capita has been widely used in empirical research, see Rodrik (1999), Acemoglu and Johnson (2014), Moretti, Steinwender, and Van Reenen (2025), and how the unemployment rate relates to GDP (or in its various forms such as lnGDP or GDP growth rate) has remained at the heart of macroeconomics and policymaking. First, we adopt a simpler version of Feng, Lagakos, and Rauch (2023) and regress the U.S. unemployment rate ($U_t$) on log GDP per capita ($\text{lnGDP}_t$) using quarterly data from July 1974 to Oct 2024, totaling 202 observations. Results are reported in Table \ref{tab:6.1}. $\hat{A}_{\hat{V}}$ denotes the OLS estimator of the prewhitening VAR(1).

\begin{center}
    [Table \ref{tab:6.1}]
\end{center}

As demonstrated earlier, regressors with nonzero mean tend to trigger eigen adjustment. Here the mean of $\text{lnGDP}_t$ is 10.7, the eigenvalues of $\hat{A}_{\hat{V}}$ are 0.883 and 0.793 in magnitude while the singular values of $\hat{A}_{\hat{V}}$ are $44.3$ and $0.016$. Therefore, eigen adjustment is triggered unnecessarily and inflates the AM-PW standard errors to be 63.6\% (21.1 versus 12.9) and 62.8\% (1.97 versus 1.21) larger than $\text{AM-PW}^{\text{unadj}}$. Depending on the estimation method, HAC standard errors differ by at most 100\% (AM-PW versus QS). Consequently, neither of the OLS coefficient is significant under AM-PW but both of them are significant at the 0.1\% level under QS. In this example, the $\text{AM-PW}^{\text{unadj}}$ standard errors are not inflated because the eigenvalues of $\hat{A}_{\hat{V}}$ are away from unity and the inverse of $I-\hat{A}_{\hat{V}}$ does not blow up.

Researchers may discover that the augmented Dickey-Fuller (ADF) tests on the quarterly data of $U_t$ and $\text{lnGDP}_t$ from July 1974 to October 2024 yield p-values of 0.049 and 0.428. Since the former marginally rejects and the latter fails to reject the presence of a unit root, alternatively both series should be differenced. Therefore, we adopt a differenced version and regress the change of the unemployment rate ($\Delta U_t=U_t-U_{t-1}$) on the GDP growth rate ($G_t=100(\text{GDP}_t/\text{GDP}_{t-1}-1)$) using quarterly data from Apr 1959 to Apr 2020, totaling 245 observations. This version is related to the Okun's law and resembles Okun's (1962) original regression with GDP replacing GNP. Results are reported in Table \ref{tab:6.2}. $\hat{B}_1$ denotes the Burg estimator of the prewhitening VAR(1).

\begin{center}
    [Table \ref{tab:6.2}]
\end{center}

Deviations in standard errors persist across the four estimators, reaching up to 140\% (QS versus $\text{AM-PW}^{\text{unadj}}$). Consequently, the intercept is significant at the 0.1\% level using AM-PW and $\text{AM-PW}^{\text{unadj}}$ but only at the 5\% level using QS and CVLL. The intercept is of particular interest to policymarkers because it has the economic interpretation that the unemployment rate will rise by $0.27\%$ from one quarter to the next if GDP does not grow.

Substantial deviations exist within the class of prewhitened kernel estimators: AM-PW, $\text{AM-PW}^{\text{unadj}}$, and CVLL, as well. Most notably, the CVLL standard errors are 126\% (0.120 versus 0.053) and 116\% (0.214 versus 0.099) larger than $\text{AM-PW}^{\text{unadj}}$. This is because the largest eigenvalue of $\hat{A}_{\hat{V}}$ has magnitude 1.46, suggesting that $\hat{V}_t = (\hat{u}_t, \hat{u}_tG_t)$ is nonstationary. Which estimator to use among the three represents orthogonal views on the DGP. AM-PW and CVLL should be used when researchers believe that $\hat{V}_t$ is stationary while $\text{AM-PW}^{\text{unadj}}$ should be used when $\hat{V}_t$ is regarded as nonstationary. A key distinction between CVLL and AM-PW is that the $\hat{B}_1$ in CVLL inherently satisfies the stationarity condition but the $\hat{A}_{\hat{V}}$ in AM-PW is manually adjusted to enforce stationarity. As the true DGP of $\hat{V}_t$ is unknown, in this example our goal is less to determine which method is correct than to emphasize that HAC standard errors are sensitive to the estimator of the prewhitening VAR.

\subsection{Purchasing Power Parity}
Purchasing power parity (PPP) describes an equilibrium in which the nominal exchange rate between two countries is equal to the price ratio of an identical bundle of goods in each country. A number of empirical tests of the relative PPP are based on the panel regression
\begin{equation*}
    \Delta S_{it} = \beta_0 + \beta_1(\pi_{it} - \pi^*_t) + u_{it},
\end{equation*}
where $\Delta S_{it}$ is the percentage change of the nominal exchange rate of country $i$'s currency relative to a foreign currency, typically the U.S. dollar, and $\pi_{it} - \pi^*_t$ is the difference of the inflation, between country $i$ and the U.S. As panel data are subject to spatial dependence that introduces additional inconsistency to standard error estimators, and since the focus of our paper is on HAC estimation, in this example we follow Froot and Rogoff (1995, p1650) and consider $i=1$, i.e., the time series regression
\begin{equation*}
    \Delta S_t = \beta_0 + \beta_1(\pi_{t} - \pi^*_t) + u_t
\end{equation*}
using annual data of UK from 1975 to 2024, totaling 50 observations. Results are reported in Table \ref{tab:6.3}. Relative PPP predicts $\beta_1=1$ so that fluctuations of the exchange rate comove with the inflation differential. Accordingly, we test whether $\hat{\beta}_1$ is significantly different from unity. For discussions on HAC estimation and PPP with spatially dependent panel data, see Driscoll and Kraay (1997).

\begin{center}
    [Table \ref{tab:6.3}]
\end{center}

Although the mean of $\pi_t - \pi^*_t$ is reduced to 1.15, the eigenvalues of $\hat{A}_{\hat{V}}$ have magnitude 0.451 and 0.268 while the singular values of $\hat{A}_{\hat{V}}$ are 1.04 and 0.116. Hence, eigen adjustment is still unnecessarily triggered and deflates the AM-PW standard error of $\hat{\beta}_1$ to be 20.9\% smaller than AM-PW$^{\text{unadj}}$. The CVLL standard error of $\hat{\beta}_1$ is the smallest among the four, with AM-PW$^{\text{unadj}}$ up to 79.9\% larger. Consequently, $\hat{\beta}_1$ is significantly different from unity at the 5\% level using CVLL but is not significantly different using the rest of the estimators. One reason for this discrepancy is that CVLL selects a prewhitening VAR of order 2 whereas QS fixes the order at 0, AM-PW and AM-PW$^{\text{unadj}}$ fix the order at 1. CVLL's choice of a higher order prewhitening VAR is supported by the fact that AIC selects VAR(16) for $\hat{V}_t = (\hat{u}_t, \hat{u}_t(\pi_{it} - \pi^*_t))$ under OLS estimation and VAR(5) under Burg estimation. Motivated by this finding, Figure 7 plots how the AM-PW$^{\text{unadj}}$ standard error of $\hat{\beta}_1$ changes with the order of the prewhitening VAR estimated by OLS without eigen adjustment.

\begin{center}
    [Figure 7]
\end{center}

Most notably, the standard error of $\hat{\beta}_1$ resulting from VAR(1) is 205\% larger than that from VAR(12), and $\hat{\beta}_1$ is significantly different from unity at the 0.1\% level using order 11 and 12 but is not significantly different using order 0, 1, 2, 3, and 14. This indicates that HAC standard errors are sensitive to the order of the prewhitening VAR and highlights the need for order selection.

\section{Conclusion}
The contributions of this paper are summarized into two parts.

On one hand, we reveal some flaws of the eigen adjustment rule of Andrews and Monahan (1992) when regressors have nonzero mean. Not only can eigen adjustment be unnecessarily triggered, but it also inflates the AM-PW standard error such that the resulting confidence interval can become too wide to be practically meaningful. Another flaw that has not been mentioned before is that HAC standard error estimators with eigen adjustment are not scale invariant. The intuition is that multiplying $y_t$ and $x_t$ by the same constant changes the mean of $\hat{u}_tx_t$, which may trigger eigen adjustment or affect its size distortion. This matters particularly when regressors are measured in different units. In the PPP example, using relative rather than percentage changes of the nominal exchange rate and CPI inflates the AM-PW standard error of $\hat{\beta}_1$ by 82\% (0.386 versus 0.212). Simply abandoning eigen adjustment and using OLS to estimate the prewhitening VAR is undesired because it can yield coefficient matrices that correspond to a nonstationary $\hat{V}_t$ when the true DGP is stationary and most HAC estimators are theoretically valid only under stationary $\hat{V}_t$. Moreover, it can produce large outliers (e.g., Table 4) and lead to substantially lower coverage rates (e.g., Table 6).

On the other hand, our initiatives of using Burg and cross-validation have improved the prewhitened kernel estimator. Estimating the prewhitening VAR with Burg guarantees the stationarity of $\hat{V}_t$ in theory and prevents outliers from occuring in simulations. Our localized FDCV method enables simultaneous order and bandwidth selection and opens the door to a more general selection of spectrum estimators. Hence, our method performs satisfactorily across all metrics: bias, MSE, and the coverage rate of the confidence interval.

The drawback of our method is computational cost, a common issue with cross-validation. Since the search region is the set of candidate spectrum estimators, in this paper CVLL$_c$ must search through every combination of order and bandwidth, and the cost further increases as sample size grows (expanding the range of bandwidths), as higher order prewhitening filters are used, and as nonparametric spectrum estimators are considered.

Asymptotic theory on CVLL$_c$ remains a priority for future work. To justify the use of CVLL$_c$, that minimizing CVLL$_c$ is asymptotically equivalent to minimizing the MISE of the spectrum estimator at zero, one could attempt to establish a localized multivariate version of equation (3.1) in Beltrao and Bloomfield (1987):
\begin{align*}
    \frac{1}{n}\text{CVLL} &= \frac{1}{n}\sum_{j=1}^{\lfloor n/2 \rfloor} \left[\log f(\omega_j) + \frac{I(\omega_j)}{f(\omega_j)}\right] + \frac{1}{2}\text{MISE} + o_p(\text{MISE}), \quad \text{where}\\
    \text{MISE} &= E\left[\frac{1}{n}\sum_{j=1}^{\lfloor n/2 \rfloor}\left(\frac{\hat{f}_{-j}(\omega_j) - f(\omega_j)}{f(\omega_j)}\right)^2\right].
\end{align*}
To establish proprieties on the order and bandwidth selected by CVLL$_c$, one could attempt to extend Theorem 5.2 in Robinson (1991). A major difficulty, however, is that Parseval's formula can no longer be used.
\newpage

\section*{References}
\refitem Acemoglu, D., \& Johnson, S. (2014). Disease and Development: a Reply to Bloom, Canning, and Fink. \textit{Journal of Political Economy}, 122(6), 1367-1375.

\refitem Andrews, D. W. K. (1991). Heteroskedasticity \& Autocorrelation Consistent Covariance Matrix Estimation. \textit{Econometrica}, 59(3), 817-858.

\refitem Andrews, D. W. K., \& Monahan, J. C. (1992). An Improved Heteroskedasticity and Autocorrelation Consistent Covariance Matrix Estimator. \textit{Econometrica}, 60(4), 953-966.

\refitem Beltrao, K. I., \& Bloomfield, P. (1987). Determining the Bandwidth of a Kernel Spectrum Estimate. \textit{Journal of Time Series Analysis}, 8(1), 21-38.

\refitem Brockwell, P. J., Dahlhaus, R., \& Trindade, A. A. (2005). Modified Burg Algorithms for Multivariate Subset Autoregression. \textit{Statistica Sinica}, 15(1), 197-213.

\refitem Burg, J. P. (1975). Maximum Entropy Spectral Analysis. Ph.D. Dissertation, Stanford University, Dept. of Geophysics.

\refitem Casini, A., \& Perron, P. (2024). Prewhitened Long-Run Variance Estimation Robust to Nonstationarity. \textit{Journal of Econometrics}, 242(1), 105794.

\refitem Den Haan, W. J., \& Levin, A. T. (2000). Robust Covariance Matrix Estimation with Data-Dependent VAR Prewhitening Order. NBER Technical Working Paper 0255.

\refitem Driscoll, J. C., \& Kraay, A. C. (1998). Consistent Covariance Matrix Estimation with Spatially Dependent Panel Data. \textit{The Review of Economics and Statistics}, 80(4), 549-560.

\refitem Fama, E. F., \& French, K. R. Common Risk Factors in the Returns on Stocks and Bonds. \textit{Journal of Financial Economics}, 33(1), 3-56.

\refitem Feng, Y., Lagakos, D., \& Rauch, J. E. (2023). Unemployment and Development. \textit{The Economic Journal}, 134(658), 614-647.

\refitem Froot, K. A., \& Rogoff, K. (1995). Perspectives on PPP and Long-Run Real Exchange Rates. \textit{The Handbook of International Economics}, 3, 1647-88.

\refitem Hurvich, M. C. (1985). Data-Driven Choice of a Spectrum Estimate: Extending the Applicability of Cross-Validation Methods. \textit{Journal of the American Statistical Association}, 80(392), 933-940.

\refitem Lee, C. C., \& Phillips, P. C. B. (1994). An ARMA Prewhitened Long-Run Variance Estimator. Manuscript, Yale University.

\refitem Liu, X., \& Chan, K. W. (2025). Positive-Definite Converging Kernel Estimation of Long-Run Variance. \textit{Journal of Business \& Economic Statistics}, 1-15.

\refitem Moretti, E., Steinwender, C., \& Van Reenen, J. (2025). The Intellectual Spoils of War? Defense R\&D, Productivity, and International Spillovers. \textit{The Review of Economics and Statistics}, 107(1), 14-27.

\refitem Morf, M., Vieria, A., \& Kailath, T. (1978). Covariance Characterization by Partial Autocorrelation Matrices. \textit{The Annals of Statistics}, 6(3), 643-648.

\refitem Newey, W. K., \& West, K. D. (1987). A Simple, Positive Semi-Definite, Heteroskedasticity and Autocorrelation Consistent Covariance Matrix. \textit{Econometrica}, 55(3), 703-708.

\refitem Newey, W. K., \& West, K. D. (1994). Automatic Lag Selection in Covariance Matrix Estimation. \textit{The Review of Economic Studies}, 61(4), 631-653.

\refitem Okun, A. M. (1962). Potential GNP: Its Measurement and Significance. In American Statistical Association, \textit{Proceedings of the Business and Economic Statistics Section}, 89-104.

\refitem Robinson, P. M. (1991). Automatic Frequency Domain Inference on Semiparametric and Nonparametric Models. \textit{Econometrica}, 59(5), 1329-1363.

\refitem Rodrik, D. (1999). Where Did All the Growth Go? External Shocks, Social Conflict, and Growth Collapses. \textit{Journal of Economic Growth}, 4(4), 385-412.

\refitem Sul D., Phillips, P. C. B., \& Choi, C.-Y. (2005). Prewhitening Bias in HAC Estimation. \textit{Oxford Bulletin of Economics and Statistics}, 67(4), 517-546.

\refitem Vats, D., \& Flegal, J. M. (2021). Lugsail Lag Windows for Estimating Time-Average Covariance Matrices. \textit{Biometrika}, 109(3), 735-750.

\refitem Xu, Z., \& Hurvich, C. M. (2023). A Unified Frequency Domain Cross-Validatory Approach to HAC Standard Error Estimation. \textit{Econometrics and Statistics}.

\section*{Appendix}
\subsection{Proof of Theorem 2.1}
We establish this proof under $E(u_t|X_0,\dots,X_{n-1}) = 0$ for $t=0,\dots,n-1$, which is much weaker than the assumptions that $\{X_t\}_{t=0}^{n-1}$ and $\{u_t\}_{t=0}^{n-1}$ are independent with $E(u_t)=0$. Let $\widetilde{X}_t = X_t - \mu_X = (0, \tilde{x}_{1t}, \dots, \tilde{x}_{dt})'$ and $||\cdot||$ denote the Euclidean norm. We verify that $\{V_t\}_{t=0}^{n-1}$ satisfies the three conditions of a zero-mean weakly stationary series.

1) $E(V_t) = E(u_tX_t) = E[E(u_tX_t|X_t)] = E[X_tE(u_t|X_t)] = 0$.

2) $E(u_t\widetilde{X}_tu_{t-r}\widetilde{X}_{t-r}') = E[u_tu_{t-r}(X_t-\mu_X)(X_{t-r}'-\mu_X')]$

$= E(u_tX_tu_{t-r}X_{t-r}') - E(u_tu_{t-r}X_t)\mu_X' - \mu_XE(u_tu_{t-r}X_{t-r}') + \gamma_u(r)\mu_X\mu_X'$. \\
$E(u_tu_{t-r}X_t) = E(u_tu_{t-r}\widetilde{X}_t) + \gamma_u(r)\mu_X$. Since $(x_{it}, u_t, u_{t-r})$ is the marginal random vector of $(x_{it}, x_{jt-r}, u_t, u_{t-r})$, $(\tilde{x}_{it}, u_t, u_{t-r})$ is a zero-mean multivariate normal random vector. By Isserlis's Theorem, $E(u_tu_{t-r}\tilde{x}_{it})=0$, then $E(u_tu_{t-r}\widetilde{X}_t) = 0$ as the first entry of $\widetilde{X}_t$ is also zero. Thus,
\begin{equation}
    \Gamma_V(r) = E(u_t\widetilde{X}_tu_{t-r}\widetilde{X}_{t-r}') + \gamma_u(r)\mu_X\mu_X'. \tag{A.1}
\end{equation}
Similarly, $(\tilde{x}_{it}, \tilde{x}_{jt-r}, u_t, u_{t-r})$ is a zero-mean multivariate normal random vector. Again by Isserlis's Theorem and the fact that the first entry of $\widetilde{X}_t$ is zero, we have
\begin{equation*}
    E(u_t\widetilde{X}_tu_{t-r}\widetilde{X}_{t-r}') = \gamma_u(r)\Gamma_X(r) + E(u_t\widetilde{X}_t)E(u_{t-r}\widetilde{X}'_{t-r}) + E(u_{t-r}\widetilde{X}_t)E(u_t\widetilde{X}_{t-r}').
\end{equation*}
Since $E(u_t|X_0,\dots,X_{n-1}) = 0$ for $t=0,\dots,n-1$, it is clear that $E(u_t\widetilde{X}_t) = E(u_{t-r}\widetilde{X}_t) = 0$. Thus, $E(u_t\widetilde{X}_t'u_{t-r}\widetilde{X}_{t-r}) = \gamma_u(r)\Gamma_X(r)$.
Combined with equation (A.1), we obtain (2.3):
\begin{equation*}
    \Gamma_V(r) = \gamma_u(r)\Gamma_X(r) + \gamma_u(r)\mu_X\mu_X'.
\end{equation*}

3) $E||V_t||^2 = E(u_t^2||X_t||^2) \leq \sqrt{E(u_t^4)E||X_t||^4} < \infty$ by Cauchy-Schwarz inequality and the assumption that $X_t$ and $u_t$ have finite fourth moments. \hfill$\square$

\subsection{Proof of Theorem 2.2}
$\hat{V}_t-V_t = X_t'(\beta-\hat{\beta})X_t$, so $||\hat{V}_t-V_t|| = |X_t'(\beta-\hat{\beta})| \ ||X_t|| \leq ||\beta-\hat{\beta}|| \ ||X_t||^2$. $||\beta-\hat{\beta}|| = o_p(1)$ follows from the assumption that $\sqrt{n}(\beta-\hat{\beta}) = O_p(1)$. $||X_t||^2 = O_p(1)$ follows from $X_t$ being weakly stationary with finite fourth moment. Hence, $\hat{V}_t-V_t = o_p(1)$. \hfill$\square$

\subsection{Proof of Theorem 3.1}
When $p=1$, expanding $V_t$ in terms of $V_{t-1}$ gives
\begin{equation*}
    V_t = AV_{t-1} + \eta_t, \quad A = \begin{bmatrix}
        \phi & 0_{1 \times d} \\
        \alpha\phi1_{d \times 1} & \phi^2I_d
    \end{bmatrix} \quad \eta_t = \begin{bmatrix}
        \tilde{\varepsilon}_t \\
        (\varepsilon_{1t}\phi u_{t-1} + \tilde{\varepsilon}_tx_{1t}, \dots, \varepsilon_{dt}\phi u_{t-1} + \tilde{\varepsilon}_tx_{dt})'
    \end{bmatrix}.
\end{equation*}
We verify that $\{\eta_t\}_{t=0}^{n-1}$ is white noise.

1) $E(\eta_t) = 0$ because $\varepsilon_{it}$ and $\tilde{\varepsilon}_t$ are independent white noises.

2) We now show that $\{\eta_t\}$ has no serial correlation, i.e., $\Gamma_\eta(r) = 0$ for $r \neq 0$.

$\Gamma_\eta(r)_{11} = E(\tilde{\varepsilon_t}\tilde{\varepsilon}_{t-r}) = 0$.

$\Gamma_\eta(r)_{1i} = E[\tilde{\varepsilon}_t(\varepsilon_{it-r}\phi u_{t-1-r} + \tilde{\varepsilon}_{t-r}x_{it-r})] = \phi E(\varepsilon_{it-r})E(\tilde{\varepsilon}_tu_{t-1-r}) + E(\tilde{\varepsilon}_t\tilde{\varepsilon}_{t-r})E(x_{it-r}) = 0$ for $i=2,\dots,n$. By symmetry, $\Gamma_\eta(r)_{i1} = E[\tilde{\varepsilon}_{t-r}(\varepsilon_{it}\phi u_{t-1} + \tilde{\varepsilon}_tx_{it})] = 0$ for $i=2,\dots,n$ follows.

$\Gamma_\eta(r)_{ij} = E[(\varepsilon_{it}\phi u_{t-1} + \tilde{\varepsilon}_tx_{it})(\varepsilon_{jt-r}\phi u_{t-1-r} + \tilde{\varepsilon}_{t-r}x_{jt-r})]$. Expanding $\Gamma_\eta(r)_{ij}$ gives
\begin{align*}
    &\phi^2E(\varepsilon_{it}\varepsilon_{jt-r})E(u_{t-1}u_{t-1-r}) +  E(\tilde{\varepsilon}_t\tilde{\varepsilon}_{t-r})E(x_{it}x_{jt-r}) \\
    + &\phi E(\varepsilon_{it}x_{jt-r})E(\tilde{\varepsilon}_{t-r}u_{t-1}) + \phi E(\varepsilon_{jt-r}x_{it})E(\tilde{\varepsilon}_tu_{t-1-r}), \quad i,j = 2,\dots,n.
\end{align*}
The first two terms are zero when $r \neq 0$. If $r>0$, $E(\varepsilon_{it}x_{jt-r}) = E(\tilde{\varepsilon}_tu_{t-1-r}) = 0$. If $r<0$, $E(\tilde{\varepsilon}_{t-r}u_{t-1}) = E(\varepsilon_{jt-r}x_{it}) = 0$. Thus, the last two terms are zero and $\Gamma_\eta(r) = 0$ for $r \neq 0$. Here we utilize $E(\varepsilon_{it}|x_{i1}, \cdots, x_{it-1}) = E(\tilde{\varepsilon}_t|u_1, \cdots, u_{t-1}) = 0$ implied by causality of $\{x_{it}\}$ and $\{u_t\}$, whose necessary and sufficient condition is the stationarity assumption.

3) By the same reasoning as in 2), one can show that $\text{Var}(\eta_t)$ is constant over time.

Because $x_{it}$ and $u_t$ are stationary AR(1) models, $|\phi| < 1$. Then the eigenvalues of $A$, $\phi$ with multiplicity 1 and $\phi^2$ with multiplicity $d$ (as evident from the expression of $A$), are within $(-1,1)$. Therefore, $V_t$ is a stationary VAR(1) model and the OLS estimator $\hat{A}$ of $A$ is consistent.

When $p \geq 2$, expanding $V_t$ in terms of $V_{t-1},\dots,V_{t-p}$ gives
\begin{align*}
    &V_t = \sum_{k=1}^pA_kV_{t-k} + \eta_t, \quad A_k = \begin{bmatrix}
        \phi_k & 0_{1 \times d} \\
        \alpha\phi_k1_{d \times 1} & \phi_k^2I_d
    \end{bmatrix} \quad \eta_t = \begin{bmatrix}
        \tilde{\varepsilon}_t \\
        (U_{1t}, \dots, U_{dt})'
    \end{bmatrix}, \\
    &\text{where} \quad U_{it} = \sum_{\substack{k,l = 1 \\ k \neq l}}^p\phi_k\phi_lu_{t-k}x_{it-l} + \varepsilon_{it}\sum_{k=1}^p\phi_ku_{t-k} + \tilde{\varepsilon}_tx_{it}, \quad i = 1,\dots,d.
\end{align*}
$\eta_t$ is autocorrelated due to the term $\sum_{\substack{k,l = 1 \\ k \neq l}}^p\phi_k\phi_lu_{t-k}x_{it-l}$, so $V_t$ is not a VAR($p$) model. By the same reasoning one can conclude that $V_t$ is not a VAR model of any (finite) order. \hfill$\square$

\subsection{Proof of Lemma 3.1}
We are left to show that the singular values of $A$ are $\phi^2$, $|s_1|$, $|s_2|$ with multiplicity $d-1$, 1, 1 respectively, where $s_1, s_2 = \frac{\phi}{2}\sqrt{\alpha^2d+(\phi+1)^2} \ \pm \ \frac{\phi}{2}\sqrt{\alpha^2d+(\phi-1)^2}$.

The singular values of $A$ are the square roots of the eigenvalues of
\begin{equation*}
    AA' = \begin{bmatrix}
        \phi^2 + \alpha^2\phi^2d & \alpha\phi^31_{1 \times d} \\
        \alpha\phi^31_{d \times 1} & \phi^4I_d
    \end{bmatrix}.
\end{equation*}
To compute the eigenvalues of $AA'$, we first consider the subspace
\begin{equation*}
    S = \{(0,u')' \in \mathbb{R}^{d+1}: u'1_{d \times 1} = 0 \ \ \text{and} \ \ u \in \mathbb{R}^d\}, \quad \dim(S) = d-1.
\end{equation*}
For any $v \in S$, $AA'v = \phi^4v$. Therefore, $\phi^4$ is an eigenvalue of $AA'$ with multiplicity $d-1$.

Next, we consider the subspace orthogonal to $S$:
\begin{equation*}
    S^\perp = \text{span}\{v_1, v_2\}, \quad v_1 = \begin{bmatrix}
        1 \\
        0_{1 \times d}
    \end{bmatrix} \quad \text{and} \quad v_2 = \begin{bmatrix}
        0 \\
        1_{1 \times d}
    \end{bmatrix}.
\end{equation*}
Since $AA'v_1 = (\phi^2 + \alpha^2\phi^2d)v_1 + \alpha\phi^3v_2$ and $AA'v_2 = \alpha\phi^3dv_1 + \phi^4v_2$, the subspace $S^\perp$ is closed under $AA'$. From standard results in linear algebra, the remaining two eigenvalues of $AA'$ are the eigenvalues of the restriction matrix
\begin{equation*}
    AA'|_{S^\perp} = \begin{bmatrix}
        \phi^2 + \alpha^2\phi^2d & \alpha\phi^3d \\
        \alpha\phi^3 & \phi^4
    \end{bmatrix}.
\end{equation*}
Calculating its eigenvalues yields $s_1^2$ and $s_2^2$, each with multiplicity 1. \hfill$\square$
\newpage

\section*{Tables and Figures}
\begin{table}[H]
    \centering
    \begin{threeparttable}
        \caption{$A = \begin{bmatrix}
            \phi & 0 \\
            \alpha\phi & \phi^2
        \end{bmatrix}$ with $\alpha = 2$}
        \begin{tabular}{|c c c c c|}
            \hline
            $\phi$ & $\{|\lambda_i(A)|\}$ & $\{\sigma_i(A)\}$ & \makecell{Triggers Eigen \\ Adjustment ?} & $\frac{||A^{\text{adj}} - A||_{1,1}}{||A||_{1,1}}$ \\
            \hline
            0.3 & \{0.3, 0.09\} & \{0.68, 0.04\} & No & 0 \\
            0.5 & \{0.5, 0.25\} & \{1.14, 0.11\} & Yes & 15.3\% \\
            0.7 & \{0.7, 0.49\} & \{1.63, 0.21\} & Yes & 41.5\% \\
            0.9 & \{0.9, 0.81\} & \{2.14, 0.34\} & Yes & 56.4\% \\
            \hline
        \end{tabular}
        \begin{tablenotes}[flushleft]
            \item[] Notes: $||A||_{1,1} = \sum_{i}\sum_{j}|A_{ij}|$. The last column measures the extent to which $A$ is modified by eigen adjustment.
        \end{tablenotes}
    \end{threeparttable}
\end{table}
\newpage

\begin{table}[H]
    \centering
    \begin{threeparttable}
        \caption{$\hat{A}_{\hat{V}}$ denotes the OLS estimator of the VAR(1) on $\hat{V}_t$ \\ with $\alpha = 2$, $n=500$, and 1000 repetitions}
        \begin{tabular}{|c c c c c|}
            \hline
            $\phi$ & $\{|\lambda_i(\hat{A}_{\hat{V}})|\}$ & $\{\sigma_i(\hat{A}_{\hat{V}})\}$ & \makecell{Frequency of \\ Eigen Adjustment} & $\frac{||\hat{A}_{\hat{V}}^{\text{adj}} - \hat{A}_{\hat{V}}||_{1,1}}{||\hat{A}_{\hat{V}}||_{1,1}}$ \\
            \hline
            0.3 & \{0.30, 0.08\} & \{0.72, 0.04\} & 21.7\% & 3.71\% \\
            0.5 & \{0.50, 0.23\} & \{1.22, 0.12\} & 63.3\% & 18.8\% \\
            0.7 & \{0.69, 0.47\} & \{1.83, 0.22\} & 83.3\% & 38.6\% \\
            0.9 & \{0.89, 0.77\} & \{3.67, 0.29\} & 96.5\% & 60.7\% \\
            \hline
        \end{tabular}
        \begin{tablenotes}[flushleft]
            \item[] Notes: $|\lambda_i(\hat{A}_{\hat{V}})|$, $\sigma_i(\hat{A}_{\hat{V}})$, and $||\hat{A}_{\hat{V}}^{\text{adj}}-\hat{A}_{\hat{V}}||_{1,1} / ||\hat{A}_{\hat{V}}||_{1,1}$ are averaged over repetitions. As a sanity check, eigen adjustment remains appropriate when $\alpha = 0$. It only happens when $\phi = 0.9$. The frequency is 1.30\% and $||\hat{A}_{\hat{V}}^{\text{adj}}-\hat{A}_{\hat{V}}||_{1,1} / ||\hat{A}_{\hat{V}}||_{1,1}$ is 0.027\%.
        \end{tablenotes}
    \end{threeparttable}
\end{table}
\newpage

\begin{figure}[H]
    \centering
    \includegraphics[width=\linewidth]{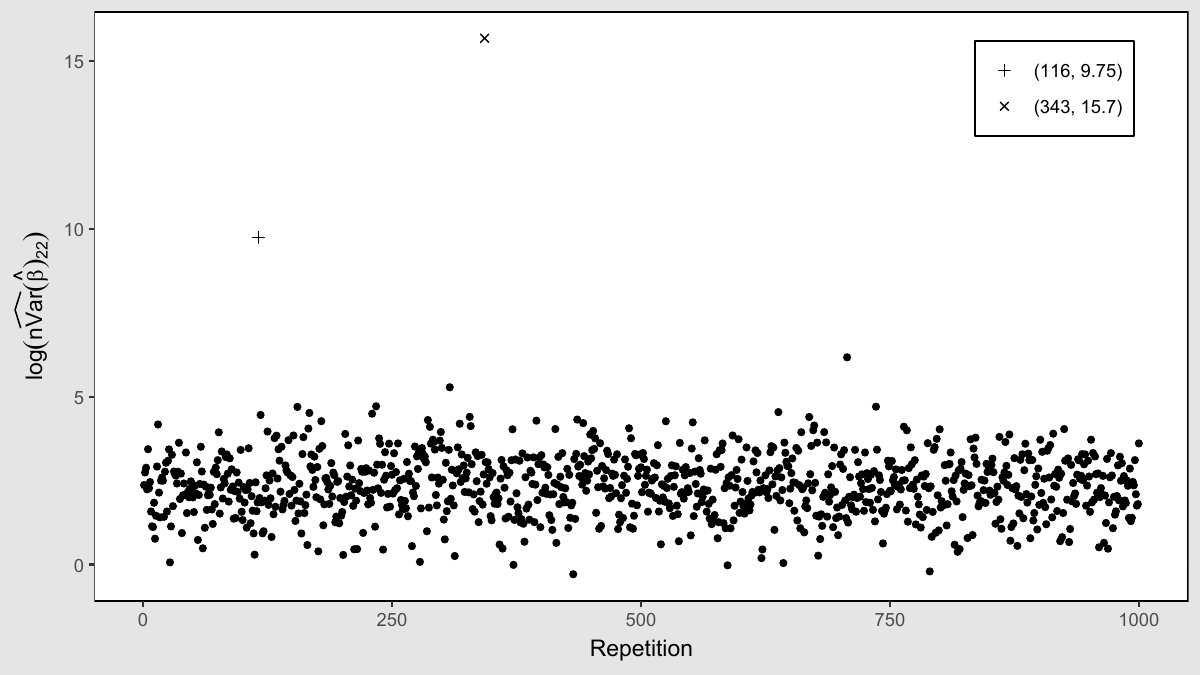}
    \caption{Log Prewhitened Kernel HAC Estimator of $n\text{Var}(\hat{\beta})_{22}$ Using \\ OLS without Eigen Adjustment to Estimate the Prewhitening VAR(1)}
\end{figure}
\begin{figure}[H]
    \centering
    \includegraphics[width=\linewidth]{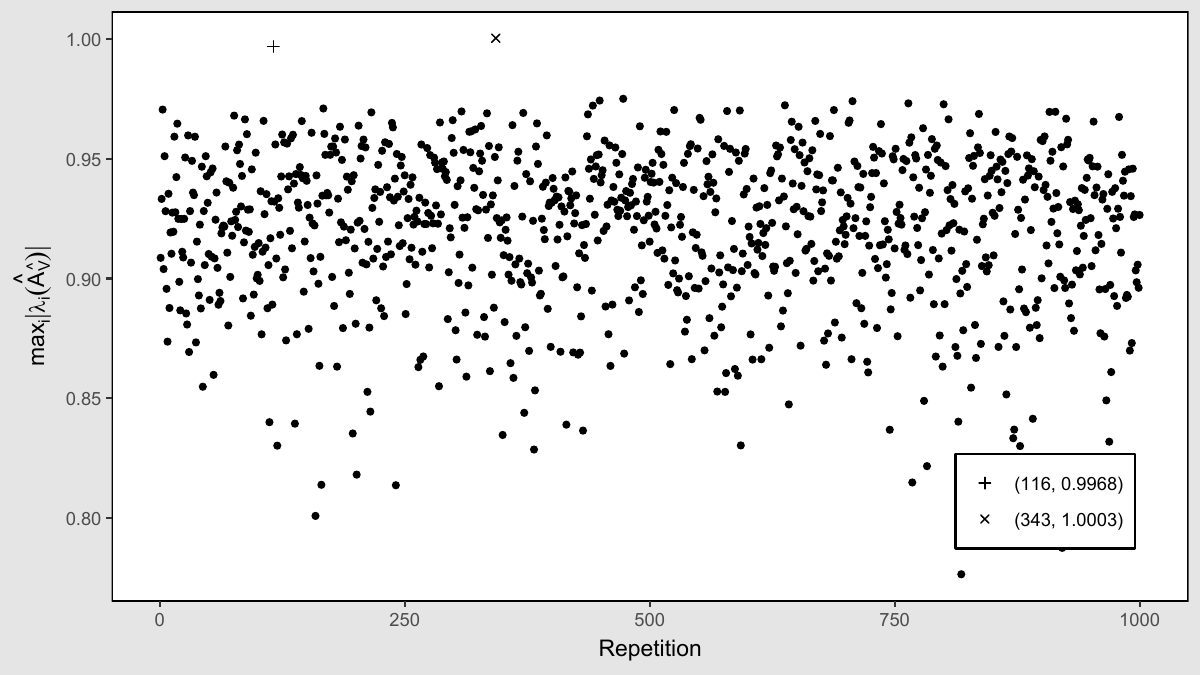}
    \caption{The Largest Eigenvalue of $\hat{A}_{\hat{V}}$ in Magnitude, \\ where $\hat{A}_{\hat{V}}$ is Estimated by OLS without Eigen Adjustment}
\end{figure}
\newpage

\begin{figure}[H]
    \centering
    \includegraphics[width=\linewidth]{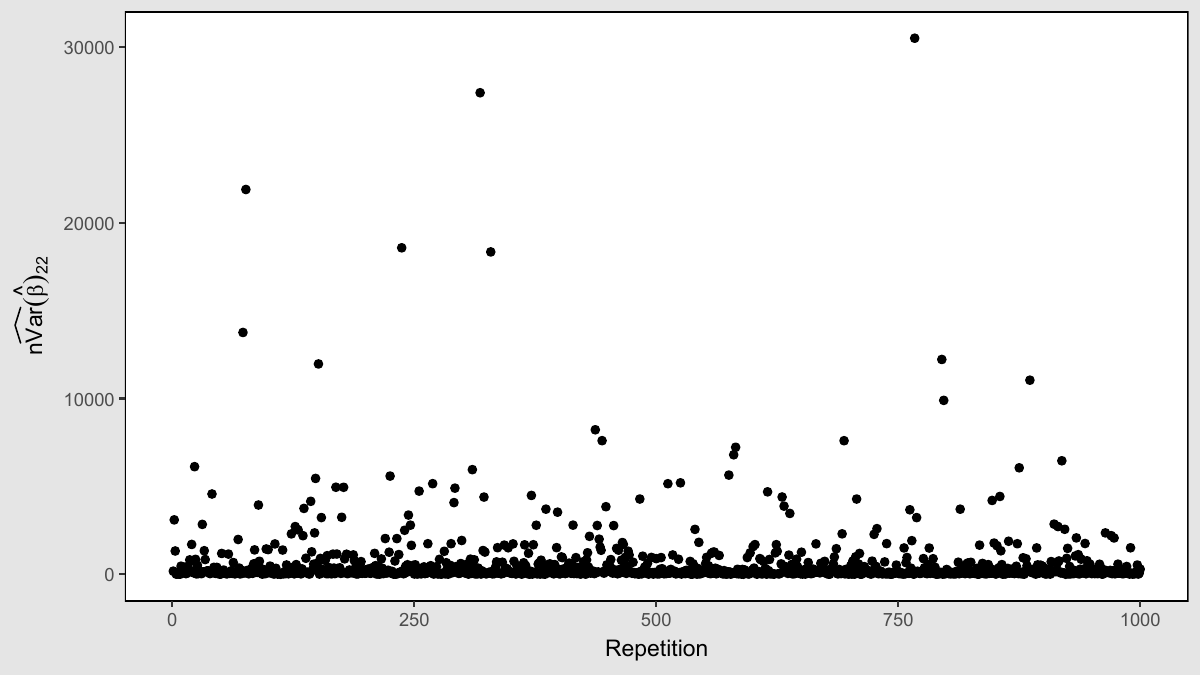}
    \caption{Prewhitened Kernel HAC Estimator of $n\text{Var}(\hat{\beta})_{22}$ Using OLS with the Eigen Adjustment Rule of Andrews and Monahan (1992) to Estimate the Prewhitening VAR(1)}
\end{figure}
\begin{figure}[H]
    \centering
    \includegraphics[width=\linewidth]{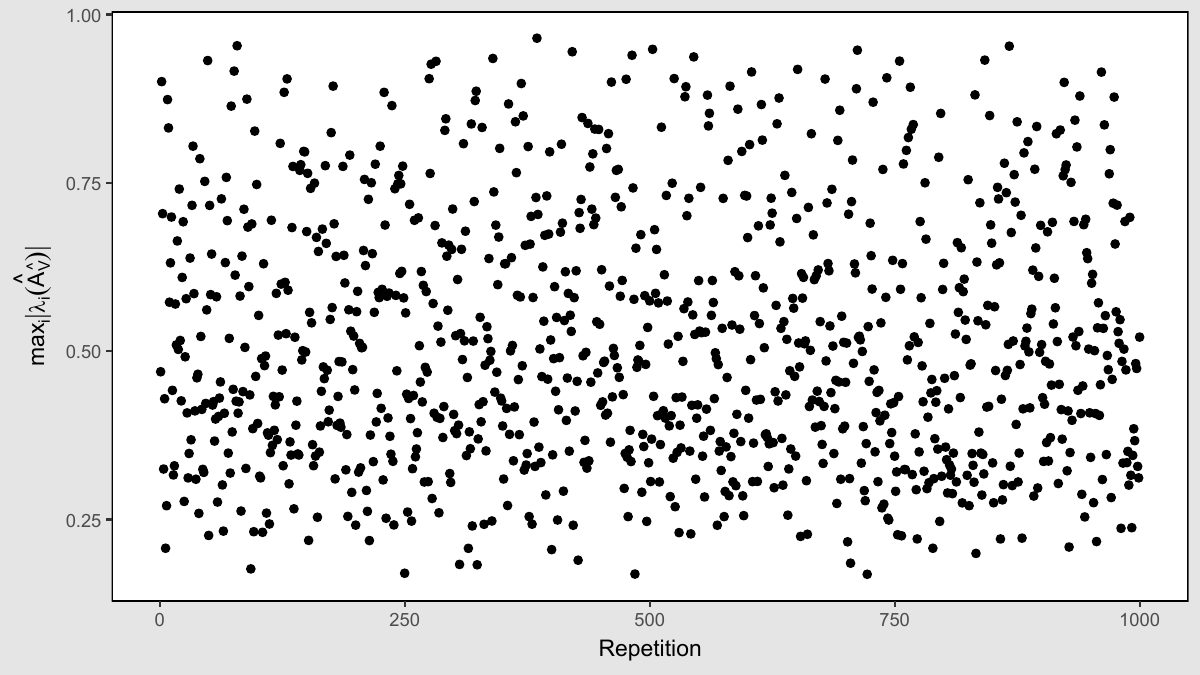}
    \caption{The Largest Eigenvalue of $\hat{A}_{\hat{V}}$ in Magnitude, where $\hat{A}_{\hat{V}}$ is Estimated by \\ OLS with the Eigen Adjustment Rule of Andrews and Monahan (1992)}
\end{figure}
\newpage

\begin{figure}[H]
    \centering
    \includegraphics[width=\linewidth]{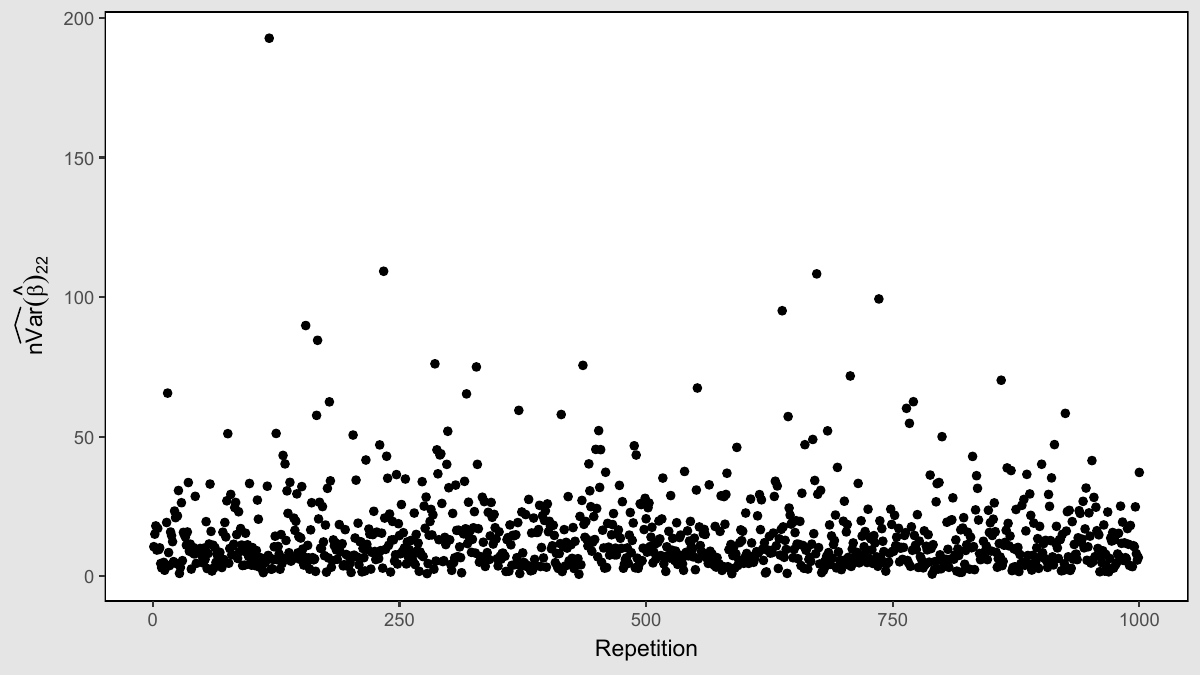}
    \caption{Prewhitened Kernel HAC Estimator of $n\text{Var}(\hat{\beta})_{22}$ Using \\ the Burg Method without Eigen Adjustment to Estimate the Prewhitening VAR(1)}
\end{figure}
\begin{figure}[H]
    \centering
    \includegraphics[width=\linewidth]{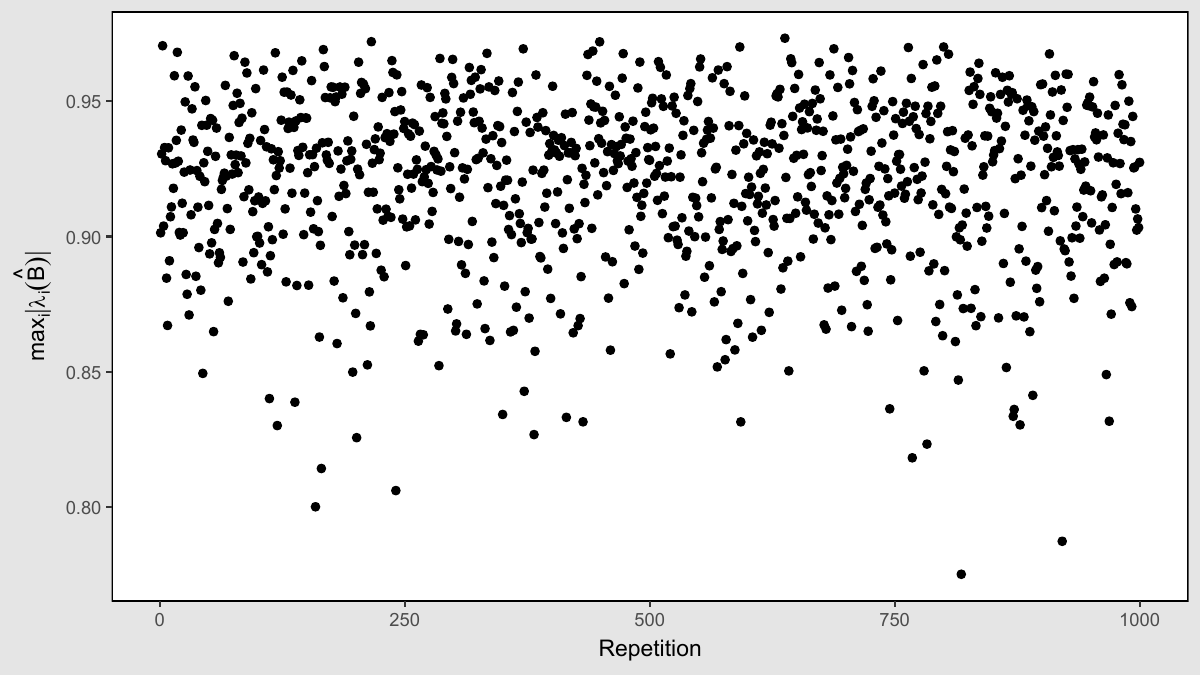}
    \caption{The Largest Eigenvalue of $\hat{B}_1$ in Magnitude, where \\ $\hat{B}_1$ is Estimated by the Burg Method without Eigen Adjustment}
\end{figure}
\newpage

\begin{table}[H]
    \centering
    \begin{threeparttable}
        \caption{Regressors and Errors are AR(1) with $n=100$ and $\alpha=0$}
        \begin{tabular}{c c c c c c c c c}
            \hline\hline
            \multirow{2}{*}{$\phi$} & \multirow{2}{*}{$n\text{Var}(\hat{\beta}_1)$} & \multirow{2}{*}{Estimator} & \multirow{2}{*}{Bias} & \multirow{2}{*}{Variance} & \multirow{2}{*}{MSE} & \multicolumn{3}{c}{Coverage Rate}\\ \cmidrule(lr){7-9}
            & & & & & & 90\% & 95\% & 99\%\\
            \hline

            0.3 & 1.16 & \makecell{QS \\ AM-PW \\ AM-PW$^\text{unadj}$ \\ CVLL} & $\makecell{-0.039 \\ \phantom{-}0.027 \\ \phantom{-}0.027 \\ -0.019}$ & \makecell{0.161 \\ 0.226 \\ 0.226 \\ 0.269} & \makecell{0.163 \\ 0.227 \\ 0.227 \\ 0.270} & \makecell{87.1 \\ 87.9 \\ 87.9 \\ 87.1} & \makecell{93.2 \\ 93.6 \\ 93.6 \\ 92.9} & \makecell{98.6 \\ 98.5 \\ 98.5 \\ 98.1}\\
            \hline

            0.6 & 2.12 & \makecell{QS \\ AM-PW \\ AM-PW$^\text{unadj}$ \\ CVLL} & $\makecell{-0.473 \\ -0.144 \\ -0.144 \\ -0.234}$ & \makecell{0.671 \\ 1.25 \\ 1.25 \\ 1.19} & \makecell{0.895 \\ 1.27 \\ 1.27 \\ 1.24} & \makecell{83.6 \\ 86.5 \\ 86.5 \\ 84.4} & \makecell{90.4 \\ 92.0 \\ 92.0 \\ 91.1} & \makecell{95.8 \\ 96.8 \\ 96.8 \\ 96.4}\\
            \hline

            0.9 & 7.89 & \makecell{QS \\ AM-PW \\ AM-PW$^\text{unadj}$ \\ CVLL} & $\makecell[l]{-4.55 \\ -2.43 \\ \phantom{-}0.287 \\ -2.56}$ & \makecell{9.20 \\ 45.0 \\ 3452 \\ 33.4} & \makecell{29.9 \\ 50.9 \\ 3452 \\ 40.0} & \makecell{65.9 \\ 74.8 \\ 75.7 \\ 74.9} & \makecell{75.3 \\ 81.4 \\ 82.3 \\ 81.9} & \makecell{84.2 \\ 89.5 \\ 90.0 \\ 89.0}\\
            \hline

            0.95 & 13.9 & \makecell{QS \\ AM-PW \\ AM-PW$^\text{unadj}$ \\ CVLL} & $\makecell[l]{-9.93 \\ -6.94 \\ \phantom{-}3.33 \\ -7.02}$ & \makecell{15.6 \\ 76.0 \\ 8259 \\ 55.6} & \makecell{114 \\ 124 \\ 8270 \\ 105} & \makecell{55.4 \\ 65.6 \\ 68.8 \\ 66.7} & \makecell{61.9 \\ 72.1 \\ 74.3 \\ 72.9} & \makecell{73.9 \\ 82.7 \\ 85.7 \\ 83.9}\\
            \hline\hline
        \end{tabular}
    \end{threeparttable}
\end{table}
\newpage

\begin{table}[H]
    \centering
    \begin{threeparttable}
        \caption{Regressors and Errors are AR(1) with $n=100$ and $\alpha=1$}
        \begin{tabular}{c c c c c c c c c}
            \hline\hline
            \multirow{2}{*}{$\phi$} & \multirow{2}{*}{$n\text{Var}(\hat{\beta}_1)$} & \multirow{2}{*}{Estimator} & \multirow{2}{*}{Bias} & \multirow{2}{*}{Variance} & \multirow{2}{*}{MSE} & \multicolumn{3}{c}{Coverage Rate}\\ \cmidrule(lr){7-9}
            & & & & & & 90\% & 95\% & 99\%\\
            \hline

            0.3 & 1.28 & \makecell{QS \\ AM-PW \\ AM-PW$^\text{unadj}$ \\ CVLL} & $\makecell{-0.170 \\ -0.109 \\ -0.108 \\ -0.144}$ & \makecell{0.172 \\ 0.212 \\ 0.218 \\ 0.259} & \makecell{0.201 \\ 0.224 \\ 0.230 \\ 0.279} & \makecell{86.0 \\ 87.3 \\ 87.2 \\ 86.3} & \makecell{91.4 \\ 91.7 \\ 91.5 \\ 91.0} & \makecell{97.0 \\ 97.3 \\ 97.3 \\ 96.8}\\
            \hline

            0.6 & 2.06 & \makecell{QS \\ AM-PW \\ AM-PW$^\text{unadj}$ \\ CVLL} & $\makecell{-0.487 \\ -0.040 \\ -0.168 \\ -0.281}$ & \makecell{0.651 \\ 3.65 \\ 1.06 \\ 0.957} & \makecell{0.889 \\ 3.66 \\ 1.09 \\ 1.04} & \makecell{81.7 \\ 84.8 \\ 85.8 \\ 84.5} & \makecell{89.8 \\ 91.6 \\ 91.9 \\ 90.7} & \makecell{95.9 \\ 97.2 \\ 97.6 \\ 96.5}\\
            \hline
            
            0.9 & 8.41 & \makecell{QS \\ AM-PW \\ AM-PW$^\text{unadj}$ \\ CVLL} & $\makecell[l]{-4.98 \\ \phantom{-}51.5 \\ \phantom{-}22.2 \\ -2.95}$ & \makecell{11.8 \\ 43087 \\ 328604 \\ 33.5} & \makecell{36.6 \\ 45739 \\ 329095 \\ 42.2} & \makecell{63.6 \\ 86.4 \\ 74.5 \\ 73.1} & \makecell{70.6 \\ 90.3 \\ 80.6 \\ 80.1} & \makecell{82.7 \\ 95.1 \\ 89.8 \\ 89.0}\\
            \hline

            0.95 & 13.2 & \makecell{QS \\ AM-PW \\ AM-PW$^\text{unadj}$ \\ CVLL} & $\makecell[l]{-9.41 \\ \phantom{-}268 \\ \phantom{-}205 \\ -6.05}$ & \makecell{15.5 \\ $4.0 \times 10^6$ \\ $3.3 \times 10^7$ \\ 60.2} & \makecell{104 \\ $4.1 \times 10^6$ \\ $3.3 \times 10^7$ \\ 96.7} & \makecell{56.5 \\ 84.0 \\ 71.3 \\ 69.2} & \makecell{63.8 \\ 88.7 \\ 76.4 \\ 74.7} & \makecell{75.1 \\ 94.7 \\ 85.6 \\ 85.1}\\
            \hline\hline
        \end{tabular}
        \begin{tablenotes}[flushleft]
            \item[] Notes: The widths of the 95\% confidence intervals for $\beta_1$ constructed using AM, AM-PW, AM-PW$^\text{unadj}$, and CVLL are 0.726, 3.03, 2.17, 0.916 when $\phi=0.9$ and 0.763, 6.57, 5.79, 1.05 when $\phi=0.95$.
        \end{tablenotes}
    \end{threeparttable}
\end{table}
\newpage

\begin{table}[H]
    \centering
    \caption{Regressors and Errors are AR(2) with $n=100$ and $\alpha=0$}
    \begin{tabular}{c c c c c c c c c}
        \hline\hline
        \multirow{2}{*}{$\phi$} & \multirow{2}{*}{$n\text{Var}(\hat{\beta}_1)$} & \multirow{2}{*}{Estimator} & \multirow{2}{*}{Bias} & \multirow{2}{*}{Variance} & \multirow{2}{*}{MSE} & \multicolumn{3}{c}{Coverage Rate}\\ \cmidrule(lr){7-9}
        & & & & & & 90\% & 95\% & 99\%\\
        \hline

        0.3 & 1.13 & \makecell{QS \\ AM-PW \\ AM-PW$^\text{unadj}$ \\ CVLL} & $\makecell{-0.176 \\ -0.163 \\ -0.163 \\ -0.129}$ & \makecell{0.124 \\ 0.156 \\ 0.156 \\ 0.238} & \makecell{0.155 \\ 0.183 \\ 0.183 \\ 0.255} & \makecell{86.1 \\ 86.4 \\ 86.4 \\ 86.0} & \makecell{92.4 \\ 92.0 \\ 92.0 \\ 91.8} & \makecell{98.1 \\ 97.5 \\ 97.5 \\ 97.0}\\
        \hline

        0.6 & 2.06 & \makecell{QS \\ AM-PW \\ AM-PW$^\text{unadj}$ \\ CVLL} & $\makecell{-0.757 \\ -0.723 \\ -0.723 \\ -0.430}$ & \makecell{0.340 \\ 0.480 \\ 0.480 \\ 0.969} & \makecell{0.973 \\ 1.00 \\ 1.00 \\ 1.15} & \makecell{77.6 \\ 78.6 \\ 78.6 \\ 81.6} & \makecell{84.5 \\ 85.0 \\ 85.0 \\ 87.4} & \makecell{94.7 \\ 94.7 \\ 94.7 \\ 95.1}\\
        \hline

        0.9 & 8.22 & \makecell{QS \\ AM-PW \\ AM-PW$^\text{unadj}$ \\ CVLL} & $\makecell[l]{-5.42 \\ -5.92 \\ -5.70 \\ -3.75}$ & \makecell{5.65 \\ 4.05 \\ 9.62 \\ 17.3} & \makecell{35.0 \\ 39.1 \\ 42.2 \\ 31.3} & \makecell{61.6 \\ 57.2 \\ 58.4 \\ 68.8} & \makecell{68.4 \\ 63.8 \\ 64.8 \\ 75.6} & \makecell{79.3 \\ 75.2 \\ 76.2 \\ 85.5}\\
        \hline

        0.95 & 12.3 & \makecell{QS \\ AM-PW \\ AM-PW$^\text{unadj}$ \\ CVLL} & $\makecell[l]{-9.00 \\ -9.75 \\ \phantom{-}12.5 \\ -6.55}$ & \makecell{11.0 \\ 9.62 \\ 454809 \\ 57.8} & \makecell{92.0 \\ 105 \\ 454966 \\ 101} & \makecell{53.0 \\ 47.0 \\ 49.0 \\ 63.6} & \makecell{62.5 \\ 55.1 \\ 56.1 \\ 70.0} & \makecell{74.5 \\ 68.1 \\ 70.7 \\ 81.9}\\
        \hline\hline
    \end{tabular}
\end{table}
\newpage

\begin{table}[H]
    \centering
    \begin{threeparttable}
        \caption{Regressors and Errors are AR(2) with $n=100$ and $\alpha=1$}
        \begin{tabular}{c c c c c c c c c}
            \hline\hline
            \multirow{2}{*}{$\phi$} & \multirow{2}{*}{$n\text{Var}(\hat{\beta}_1)$} & \multirow{2}{*}{Estimator} & \multirow{2}{*}{Bias} & \multirow{2}{*}{Variance} & \multirow{2}{*}{MSE} & \multicolumn{3}{c}{Coverage Rate}\\ \cmidrule(lr){7-9}
            & & & & & & 90\% & 95\% & 99\%\\
            \hline
    
            0.3 & 1.22 & \makecell{QS \\ AM-PW \\ AM-PW$^\text{unadj}$ \\ CVLL} & $\makecell{-0.162 \\ -0.146 \\ -0.150 \\ -0.127}$ & \makecell{0.147 \\ 0.167 \\ 0.174 \\ 0.264} & \makecell{0.173 \\ 0.188 \\ 0.197 \\ 0.280} & \makecell{86.5 \\ 85.8 \\ 85.8 \\ 86.1} & \makecell{92.2 \\ 92.5 \\ 92.4 \\ 92.3} & \makecell{97.4 \\ 97.6 \\ 97.6 \\ 97.5}\\
            \hline
    
            0.6 & 1.88 & \makecell{QS \\ AM-PW \\ AM-PW$^\text{unadj}$ \\ CVLL} & $\makecell{-0.504 \\ -0.576 \\ -0.552 \\ -0.234}$ & \makecell{0.463 \\ 0.451 \\ 0.387 \\ 0.937} & \makecell{0.717 \\ 0.783 \\ 0.692 \\ 0.991} & \makecell{80.9 \\ 79.8 \\ 79.6 \\ 83.8} & \makecell{88.5 \\ 87.6 \\ 87.5 \\ 90.5} & \makecell{95.6 \\ 94.9 \\ 95.5 \\ 96.3}\\
            \hline
    
            0.9 & 8.22 & \makecell{QS \\ AM-PW \\ AM-PW$^\text{unadj}$ \\ CVLL} & $\makecell[l]{-5.35 \\ \phantom{-}8.28 \\ -5.79 \\ -3.85}$ & \makecell{5.67 \\ 10984 \\ 8.68 \\ 17.5} & \makecell{34.3 \\ 11052 \\ 42.2 \\ 32.4} & \makecell{63.7 \\ 65.4 \\ 59.5 \\ 71.4} & \makecell{70.9 \\ 72.8 \\ 67.2 \\ 76.9} & \makecell{81.3 \\ 81.9 \\ 77.6 \\ 85.8}\\
            \hline
                
            0.95 & 13.4 & \makecell{QS \\ AM-PW \\ AM-PW$^\text{unadj}$ \\ CVLL} & $\makecell[l]{-10.3 \\ \phantom{-}75.6 \\ -10.3 \\ -8.22}$ & \makecell{10.0 \\ $1.2 \times 10^6$ \\ 93.9 \\ 35.3} & \makecell{116 \\ $1.2 \times 10^6$ \\ 200 \\ 103} & \makecell{51.8 \\ 59.8 \\ 47.5 \\ 60.3} & \makecell{58.9 \\ 66.5 \\ 55.9 \\ 67.7} & \makecell{70.6 \\ 74.6 \\ 67.2 \\ 77.9}\\
            \hline\hline
        \end{tabular}
         \begin{tablenotes}[flushleft]
            \item[] Notes: The widths of the 95\% confidence intervals for $\beta_1$ constructed using AM, AM-PW, AM-PW$^\text{unadj}$, and CVLL are 0.664, 1.59, 0.610, 0.819 when $\phi=0.9$ and 0.698, 3.70, 0.695, 0.896 when $\phi=0.95$.
        \end{tablenotes}
    \end{threeparttable}
\end{table}
\newpage

\begin{table}[H]
    \centering
    \begin{threeparttable}
        \caption{Regressors and Errors are AR(3) with $n=200$ and $\alpha=0$}
        \begin{tabular}{c c c c c c c c c}
            \hline\hline
            \multirow{2}{*}{$\phi$} & \multirow{2}{*}{$n\text{Var}(\hat{\beta}_1)$} & \multirow{2}{*}{Estimator} & \multirow{2}{*}{Bias} & \multirow{2}{*}{Variance} & \multirow{2}{*}{MSE} & \multicolumn{3}{c}{Coverage Rate} \\ \cmidrule(lr){7-9}
            & & & & & & 90\% & 95\% & 99\% \\
            \hline
    
            0.3 & 1.12 & \makecell{QS \\ AM-PW \\ AM-PW$^\text{unadj}$ \\ CVLL} & $\makecell{-0.113 \\ -0.104 \\ -0.104 \\ -0.090}$ & \makecell{0.054 \\ 0.069 \\ 0.069 \\ 0.099} & \makecell{0.067 \\ 0.079 \\ 0.079 \\ 0.107} & \makecell{87.2 \\ 86.7 \\ 86.7 \\ 86.7} & \makecell{93.5 \\ 93.3 \\ 93.3 \\ 93.0} & \makecell{98.3 \\ 98.2 \\ 98.2 \\ 98.2}\\
            \hline
    
            0.6 & 1.80 & \makecell{QS \\ AM-PW \\ AM-PW$^\text{unadj}$ \\ CVLL} & $\makecell{-0.610 \\ -0.600 \\ -0.600 \\ -0.396}$ & \makecell{0.143 \\ 0.144 \\ 0.144 \\ 0.315} & \makecell{0.515 \\ 0.503 \\ 0.503 \\ 0.472} & \makecell{80.7 \\ 81.1 \\ 81.1 \\ 83.8} & \makecell{88.7 \\ 88.9 \\ 88.9 \\ 91.1} & \makecell{95.5 \\ 95.9 \\ 95.9 \\ 97.2}\\
            \hline
    
            0.9 & 10.7 & \makecell{QS \\ AM-PW \\ AM-PW$^\text{unadj}$ \\ CVLL} & $\makecell[l]{-7.58 \\ -8.76 \\ -8.74 \\ -6.96}$ & \makecell{4.90 \\ 1.42 \\ 1.54 \\ 9.18} & \makecell{62.3 \\ 78.2 \\ 77.9 \\ 57.6} & \makecell{58.8 \\ 48.4 \\ 48.5 \\ 61.9} & \makecell{66.6 \\ 56.9 \\ 56.9 \\ 70.7} & \makecell{79.5 \\ 69.4 \\ 69.6 \\ 82.3}\\
            \hline
                
            0.95 & 17.3 & \makecell{QS \\ AM-PW \\ AM-PW$^\text{unadj}$ \\ CVLL} & $\makecell[l]{-13.0 \\ -15.1 \\ -15.0 \\ -12.4}$ & \makecell{15.8 \\ 3.04 \\ 5.02 \\ 27.6} & \makecell{186 \\ 232 \\ 229 \\ 180} & \makecell{53.3 \\ 41.2 \\ 42.3 \\ 55.0} & \makecell{61.3 \\ 48.4 \\ 49.6 \\ 63.4} & \makecell{74.9 \\ 59.5 \\ 61.2 \\ 76.7}\\
            \hline\hline
        \end{tabular}
    \end{threeparttable}
\end{table}
\newpage

\begin{table}[H]
    \centering
    \begin{threeparttable}
        \caption{Regressors and Errors are AR(3) with $n=200$ and $\alpha=1$}
        \begin{tabular}{c c c c c c c c c}
            \hline\hline
            \multirow{2}{*}{$\phi$} & \multirow{2}{*}{$n\text{Var}(\hat{\beta}_1)$} & \multirow{2}{*}{Estimator} & \multirow{2}{*}{Bias} & \multirow{2}{*}{Variance} & \multirow{2}{*}{MSE} & \multicolumn{3}{c}{Coverage Rate} \\ \cmidrule(lr){7-9}
            & & & & & & 90\% & 95\% & 99\% \\
            \hline
    
            0.3 & 1.15 & \makecell{QS \\ AM-PW \\ AM-PW$^\text{unadj}$ \\ CVLL} & $\makecell{-0.125 \\ -0.114 \\ -0.113 \\ -0.108}$ & \makecell{0.057 \\ 0.067 \\ 0.068 \\ 0.102} & \makecell{0.072 \\ 0.080 \\ 0.081 \\ 0.114} & \makecell{87.5 \\ 87.8 \\ 87.8 \\ 87.4} & \makecell{93.1 \\ 93.8 \\ 93.8 \\ 92.8} & \makecell{98.1 \\ 98.3 \\ 98.3 \\ 98.0}\\
            \hline
    
            0.6 & 1.95 & \makecell{QS \\ AM-PW \\ AM-PW$^\text{unadj}$ \\ CVLL} & $\makecell{-0.633 \\ -0.766 \\ -0.745 \\ -0.540}$ & \makecell{0.216 \\ 0.146 \\ 0.138 \\ 0.309} & \makecell{0.616 \\ 0.732 \\ 0.693 \\ 0.601} & \makecell{81.3 \\ 78.5 \\ 79.0 \\ 81.9} & \makecell{88.2 \\ 87.0 \\ 87.1 \\ 88.9} & \makecell{95.0 \\ 94.7 \\ 94.9 \\ 95.7}\\
            \hline
    
            0.9 & 10.2 & \makecell{QS \\ AM-PW \\ AM-PW$^\text{unadj}$ \\ CVLL} & $\makecell[l]{-6.65 \\ \phantom{-}1.22 \\ -8.18 \\ -6.59}$ & \makecell{7.54 \\ 1442 \\ 1.76 \\ 9.15} & \makecell{51.7 \\ 1443 \\ 68.6 \\ 52.5} & \makecell{60.9 \\ 64.4 \\ 51.2 \\ 61.4} & \makecell{70.4 \\ 70.1 \\ 57.6 \\ 70.3} & \makecell{82.2 \\ 80.9 \\ 71.2 \\ 82.0}\\
            \hline
                
            0.95 & 17.3 & \makecell{QS \\ AM-PW \\ AM-PW$^\text{unadj}$ \\ CVLL} & $\makecell[l]{-12.3 \\ \phantom{-}93.6 \\ -14.8 \\ -12.0}$ & \makecell{29.1 \\ $1.6 \times 10^6$ \\ 6.84 \\ 58.5} & \makecell{181 \\ $1.6 \times 10^6$ \\ 226 \\ 202} & \makecell{55.7 \\ 58.1 \\ 42.1 \\ 55.9} & \makecell{63.4 \\ 64.0 \\ 49.4 \\ 64.1} & \makecell{76.6 \\ 74.5 \\ 62.2 \\ 76.1}\\
            \hline\hline
        \end{tabular}
    \end{threeparttable}
\end{table}
\newpage

\begin{table}[H]
    \centering
    \begin{threeparttable}
        \caption{Regressors and Errors are MA(2) with $n=200$ and $\alpha=0$}
        \begin{tabular}{c c c c c c c c c}
            \hline\hline
            \multirow{2}{*}{$\{\phi_1, \phi_2\}$} & \multirow{2}{*}{$n\text{Var}(\hat{\beta}_1)$} & \multirow{2}{*}{Estimator} & \multirow{2}{*}{Bias} & \multirow{2}{*}{Variance} & \multirow{2}{*}{MSE} & \multicolumn{3}{c}{Coverage Rate} \\ \cmidrule(lr){7-9}
            & & & & & & 90\% & 95\% & 99\% \\
            \hline
    
            $\{0,0.6\}$ & 1.39 & \makecell{QS \\ AM-PW \\ AM-PW$^\text{unadj}$ \\ CVLL} & $\makecell{-0.363 \\ -0.384 \\ -0.384 \\ -0.099}$ & \makecell{0.069 \\ 0.074 \\ 0.074 \\ 0.254} & \makecell{0.201 \\ 0.221 \\ 0.221 \\ 0.264} & \makecell{84.0 \\ 83.0 \\ 83.0 \\ 89.1} & \makecell{89.7 \\ 89.4 \\ 89.4 \\ 94.1} & \makecell{96.8 \\ 96.7 \\ 96.7 \\ 98.6}\\
            \hline

            $\{0,-0.6\}$ & 1.40 & \makecell{QS \\ AM-PW \\ AM-PW$^\text{unadj}$ \\ CVLL} & $\makecell{-0.348 \\ -0.368 \\ -0.368 \\ -0.092}$ & \makecell{0.080 \\ 0.089 \\ 0.089 \\ 0.279} & \makecell{0.201 \\ 0.225 \\ 0.225 \\ 0.288} & \makecell{83.3 \\ 82.6 \\ 82.6 \\ 90.5} & \makecell{90.1 \\ 89.8 \\ 89.8 \\ 94.3} & \makecell{96.8 \\ 96.7 \\ 96.7 \\ 98.4}\\
            \hline
            
            $\{0.3,0.6\}$ & 1.70 & \makecell{QS \\ AM-PW \\ AM-PW$^\text{unadj}$ \\ CVLL} & $\makecell{-0.423 \\ -0.477 \\ -0.477 \\ -0.049}$ & \makecell{0.156 \\ 0.142 \\ 0.142 \\ 0.408} & \makecell{0.341 \\ 0.369 \\ 0.369 \\ 0.410} & \makecell{83.1 \\ 82.3 \\ 82.3 \\ 87.4} & \makecell{89.5 \\ 89.0 \\ 89.0 \\ 93.5} & \makecell{96.4 \\ 96.4 \\ 96.4 \\ 98.0}\\
            \hline
                
            $\{0.3,-0.6\}$ & 1.41 & \makecell{QS \\ AM-PW \\ AM-PW$^\text{unadj}$ \\ CVLL} & $\makecell{-0.358 \\ -0.380 \\ -0.380 \\ -0.059}$ & \makecell{0.072 \\ 0.075 \\ 0.075 \\ 0.263} & \makecell{0.200 \\ 0.219 \\ 0.219 \\ 0.266} & \makecell{84.2 \\ 83.1 \\ 83.1 \\ 87.5} & \makecell{89.3 \\ 89.2 \\ 89.2 \\ 93.4} & \makecell{96.8 \\ 97.1\\ 97.1 \\ 98.9}\\
            \hline\hline
        \end{tabular}
    \end{threeparttable}
\end{table}
\newpage

\begin{table}[H]
    \centering
    \begin{threeparttable}
        \caption{$U_t = \beta_0 + \beta_1\text{lnGDP} + u_t$}
        \begin{tabular}{|c c c c c c|}
            \hline
            & \multirow{2}{*}{\makecell{OLS \\ Coefficient}} & \multicolumn{4}{c|}{Standard Error Estimator} \\ \cline{3-6}
            & & QS & AM-PW & $\text{AM-PW}^{\text{unadj}}$ & CVLL \\
            \hline
            $\hat{\beta}_0$ & \phantom{$-$}39.6 & $10.5^{***}$ & 21.1 & $12.9^{**}$ & $15.2^{**}$ \\
            $\hat{\beta}_1$ & $-3.11$ & $0.993^{***}$ & 1.97 & $\!\!1.21^*$ & $\!\!1.41^*$ \\
            \hline
        \end{tabular}
        \begin{tablenotes}[flushleft]
            \item[] Notes: The unemployment rate and GDP data are drawn from the U.S. Bureau of Labor Statistics and the Federal Reserve Bank of St. Louis. We use real instead of nominal GDP per capita. * $p<0.05$, ** $p<0.01$, *** $p<0.001$.
        \end{tablenotes}
        \label{tab:6.1}
    \end{threeparttable}
\end{table}
\begin{table}[H]
    \centering
    \caption{$\Delta U_t = \beta_0 + \beta_1G_t + u_t$}
    \begin{tabular}{|c c c c c c|}
        \hline
        & \multirow{2}{*}{\makecell{OLS \\ Coefficient}} & \multicolumn{4}{c|}{Standard Error Estimator} \\ \cline{3-6}
        & & QS & AM-PW & $\text{AM-PW}^{\text{unadj}}$ & CVLL \\
        \hline
        $\hat{\beta}_0$ & \phantom{$-$}0.291 & $0.128^*$ & $0.079^{***}$ & $0.053^{***}$ & $\!\!0.120^{*}$ \\
        $\hat{\beta}_1$ & $-0.559$ & $0.228^{*}$ & $0.144^{***}$ & $0.099^{***}$ & $0.214^{**}$ \\
        \hline
    \end{tabular}
    \label{tab:6.2}
\end{table}
\newpage

\begin{table}[H]
    \centering
    \begin{threeparttable}
        \caption{$\Delta S_t = \beta_0 + \beta_1(\pi_t-\pi^*_t) + u_t$}
        \begin{tabular}{|c c c c c c|}
            \hline
            & \multirow{2}{*}{\makecell{OLS \\ Coefficient}} & \multicolumn{4}{c|}{Standard Error Estimator} \\ \cline{3-6}
            & & QS & AM-PW & $\text{AM-PW}^{\text{unadj}}$ & CVLL \\
            \hline
            $\hat{\beta}_0$ & 0.768 & 1.39 & 0.929 & 1.58 & 1.28 \\
            $\hat{\beta}_1$ & 0.698 & 0.265 & 0.212 & 0.268 & $0.149^{*}$ \\
            \hline
        \end{tabular}
        \begin{tablenotes}[flushleft]
            \item[] Notes: The nominal exchange rate and inflation data are drawn from the World Bank Group. Inflation is measured as the percentage change of the consumer price index (CPI). Alternatively, $\Delta S_t$ and $\pi_t$ can be measured as the log difference of the nominal exchange rate and CPI.
        \end{tablenotes}
        \label{tab:6.3}
    \end{threeparttable}
\end{table}
\begin{figure}[H]
    \centering
    \includegraphics[width=\linewidth]{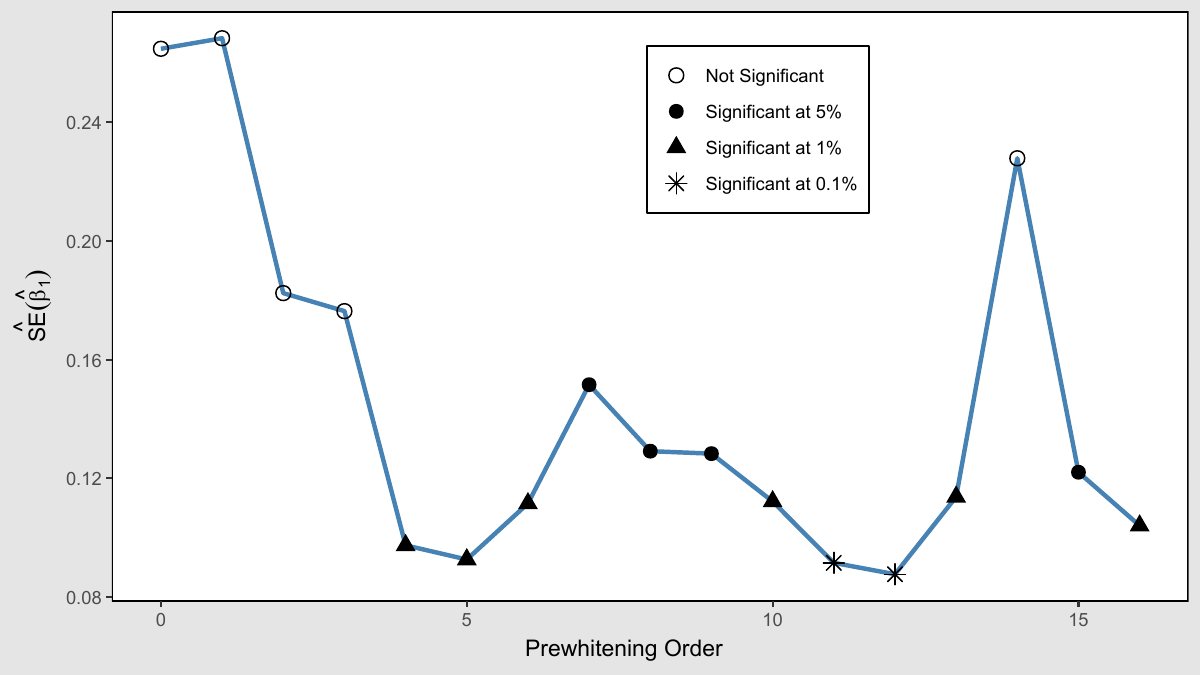}
    \caption{The AM-PW$^\text{unadj}$ Standard Error of $\hat{\beta}_1$ Using Prewhitening VAR of Order 0 to 16}
\end{figure}
\end{document}